\documentclass[twoside,11pt]{article}
\usepackage{jmlr2e}
\setkeys{Gin}{clip=true,draft=false}
\DeclareGraphicsExtensions{.pdf}
\usepackage{amsmath}
\usepackage{amsfonts}
\usepackage{algorithm}
\usepackage[noend]{algpseudocode}
\usepackage{caption}
\usepackage{subcaption}
\usepackage{color}
\usepackage{braket}
\usepackage{yfonts}
\usepackage[inline]{enumitem}
\usepackage{centernot}
\newcommand{\mbf}{\mathbf}
\newcommand{\bs}{\boldsymbol}

\newcommand{\E}{\mathbf{E}}
\newcommand{\Var}{\text{Var}}

\newcommand{\bsta}{\bs{\theta}}
\newcommand{\bsph}{\bs{\phi}}

\begin{document}

\begin{titlepage}

\title{
The Optimal Deterrence of Crime: \\
A Focus on the Time Preference of DWI Offenders
}

\maketitle

\begin{center}
Yuqing Wang$^1$ \quad and \quad Yan Ru Pei$^{2*}$ \\
\end{center}

\begin{center}
{\it 1. Department of Sociology, University of Macau \\ Avenida da Universidade, Taipa, Macau, China} \\
email: {\it yuqing.wang@connect.um.edu.mo \\}
\end{center}

\begin{center}
{\it 2. Department of Physics, University of California \\ 
San Diego, 9500 Gilman Drive, La Jolla, California 92093-0319, USA} \\
email: {\it yrpei@ucsd.edu }
\end{center}

* corresponding author

\vspace{20px}

Declaration of interest: none

\clearpage
\pagenumbering{arabic} 
\end{titlepage}

\begin{abstract}

We develop a general model for finding the optimal {\it penal strategy} based on the behavioral traits of the offenders. We focus on how the discount rate (level of time discounting) affects criminal propensity on the individual level, and how the aggregation of these effects influences criminal activities on the population level. The effects are aggregated based on the distribution of discount rate among the population. We study this distribution empirically through a survey with 207 participants, and we show that it follows zero-inflated exponential distribution. We quantify the effectiveness of the penal strategy as its net utility for the population, and show how this quantity can be maximized. When we apply the maximization procedure on the offense of impaired driving (DWI), we discover that the effectiveness of DWI deterrence depends critically on the amount of fine and prison condition.

\end{abstract}


\begin{jel}
C12, C25, C61, C83, I31, K14, K42
\end{jel}

\section{Introduction}

It is commonly accepted that crimes are detrimental to the general well-being of a society, so it is socially favorable to reduce the level of criminal activities through the implementation of certain legal practices \citep{crime}. In some modern theories of crime, this argument is formalized using an economic model, which attempts to characterize a criminal offense by the social disutility it generates for the population \citep{becker}. The net disutility is evaluated as the difference of the benefit it generates for the offender and the cost it incurs for the victim. And by taking the sum of the disutilities of all criminal offenses, we obtain the total disutility of criminal activities in the population. Therefore, through reducing the level of criminal activities, the total disutility is effectively decreased, thus achieving a more socially favorable state \citep{crime_sowel} for the population. \\

It is important to predict the effectiveness of the legal practices before they are implemented. To do so accurately, it is necessary to first predict the level of criminal activities in a population. This is done through an understanding of the underlying mechanism behind how the level of criminal activities is influenced by certain legal practices. This understanding can be achieved by applying the concept of deterrence in penology, which states that the threat of punishment prevents a person from committing a crime \citep{deter}. We can formalize this concept under the framework of rational choice theory \citep{utility} by focusing on the net utility of crime for the offender. The net utility is simply the difference of the utility that the crime rewards the offender and the expected disutility that the potential punishment incurs. Therefore, by increasing the threat of punishment, the expected disutility of punishment will be greater, and the offender will be less inclined to commit the crime. If the threat of punishment increases to the point where the net utility becomes negative for the offender, then the offender will decide against committing the crime (in accordance with rational choice theory), and a successful deterrence is achieved. To quantify the ``threat of punishment" as a disutility, we denote the specific implementation of the legal practices as a {\it penal strategy}, which we specify with factors such as the severity, certainty, and celerity of the punishment \citep{deter}. \\

A penal strategy generates social benefits as the reduction of criminal activities; however, it also generates social costs as a direct result of its implementation \citep{cost_imp}. These costs generally increase with the threat of punishment that the strategy imposes. For example, increasing the certainty of punishment implies using more police resources for criminal detection, and increasing the severity of punishment implies using more prison resources to enforce a longer term of imprisonment. Besides the implementation cost, the punishment itself is also a disutility for the offender, which appears as the opportunity cost, stigma, and any dissatisfaction that is associated with the punishment \citep{punish}. It is crucial that this disutility is included as a cost of the penal strategy, as the utility function of the offender is also part of the utility function of the society \citep{sowel}. The optimal penal strategy should aim to maximize the difference of the all benefits and costs it generates. \\

The social benefit of the penal strategy is associated with the {\it nonoccurrence} of crimes while the social cost is associated with the {\it occurrence} (and the ensuing punishment) of crimes. To evaluate the benefits and costs, it is necessary to predict the number of criminal offenses under the penal strategy. Before we make this prediction on the population level, we first study the occurrence of a criminal offense under the penal strategy on the individual level. To begin with, we assume that whether a crime occurs or not depends on various traits of the offender and victim \citep{behave}, which we can denote as the {\it parameterization} of their utility functions, or {\it utility parameters}. Individuals with different utility parameters evaluate the net utility of the crime differently, resulting in the difference in their decisions of whether to commit the crime or not. For example, offenders with different perceptions of risk may evaluate the disutility of the threat of punishment differently, and those who perceive the disutility to be greater are less likely to commit the crime. Therefore, it is possible to make an {\it a priori} categorization of an individual as a potential offender or non-offender based on how he/she evaluates the net utility of crime using his/her utility parameters. We then extend this analysis to the population level by studying how these parameters are {\it distributed} among the population. We first evaluate the net utility of crime for every individual under the penal strategy in effect, and record the number of times that the net utilities evaluate to a negative value. This number is then equal to the predicted number of criminal offenses in the population, which we can use to compute the benefits and costs associated with the penal strategy. The difference of benefits and costs is the net utility of the penal strategy for the population, and the goal is to find the optimal penal strategy that maximizes this net benefit. \\

As an example, we focus on the time preference of an individual as his/her major behavioral trait, and study how this trait influences his/her criminal propensity towards committing the offense of impaired driving (DWI). In this work, an individual's time preference is described using the hyperbolic time discounting model \citep{hyper}, under which the level of time preference can be quantified as the rate at which the discounting function decays, or {\it discount rate}. In short, a person with a larger discount rate will assign a higher value to present rewards relative to future rewards, so he/she is more likely going to commit a crime as the relative value of the immediate reward from the crime is greater with respect to the potential future punishment \citep{crime_hyper}. The discount rate then serves as a strong predictor of whether an individual will become a DWI offender, and by knowing how the discount rate is distributed among a population, we can accurately predict the number of DWI incidents under a penal strategy. The goal of this paper is then to empirically measure this distribution, evaluate explicitly the net social benefit that a penal strategy generates based on this distribution, and find the penal strategy that maximizes this net social benefit. \\

We first briefly present the structure of the paper. In section \ref{pre}, we describe the time preference and risk perception of an individual in terms of {\it hyperbolic time discounting} \citep{hyper} and {\it probability weighing function} \citep{prospect}, and discuss how these two traits generate variations in individual criminal propensities. In section \ref{con_opt}, we use a graph theoretical \citep{graph} approach to model the criminal activities in a population as a directed graph, with an arrow denoting a criminal offense pointing from the offender to the victim. Given a penal strategy, the graph is partitioned into two subgraphs, one consisting of the offenders and one consisting of the non-offenders\footnote{This is technically not the case as we are partitioning the edges instead of the nodes. However, in section \ref{dwi_case}, we show that the two partitioning schemes are equivalent in the context of DWI.}. The {\it social welfare function} \citep{sowel} is constructed as the sum over the graph connections. In section \ref{dwi_case}, we derive an explicit expression for the social welfare function based on several simplifying assumptions in the context of DWI. In section \ref{solve_opt}, we find the optimal penal strategy by maximizing the social welfare function, and show that abrupt changes in the optimum can be realized by varying the amount of fine and prison conditions. In section \ref{emp}, we estimate empirically how the {\it discount rate} (denoting the level of time preference) and {\it probability weighing factor} (denoting the level of risk aversion) are distributed among the population by conducting a survey for 207 participants; in addition, we show that the two behavioral traits are distributed independently. In section \ref{ext}, we discuss several possible extensions to our model which can be made to account for a wider range of behavioral traits and more complex criminal patterns. 

\section{Preliminaries}
\label{pre}

The main goal of this paper is to find the {\it optimal penal strategy}. However, at this point, this is an ill-defined goal as we have not yet defined exactly what constitutes a {\it penal strategy} (or what we are trying to optimize over) and exactly what this penal strategy should aim to maximize (or what we are trying to find the optimum of). To formally define the optimization problem, we first have to introduce the necessary concepts in penology and economy. We first provide a brief overview of the concept of {\it deterrence} in criminology \citep{deter}, and introduce the object of {\it penal strategy} which quantifies the ``level" of deterrence. We then briefly discuss the concepts of hyperbolic time discounting \citep{hyper} and prospect theory \citep{prospect} in the context of deterrence. Finally, we illustrate how a person's behavioral traits can be specified by the {\it parameters} of his/her utility function \citep{utility}, and we use a person's {\it discount rate} as an illustrative example.

\subsection{Penal Strategy}
\label{det_sec}

Deterrence theory states that the threat of punishment prevents people from committing crimes. In its traditional formulation, a penal strategy is specified by three main factors - the severity, certainty, and celerity of punishment. and the celerity of punishment \citep{deter}. It is generally believed that the strength of each factor is correlated with the level of criminal activities in a population, and there has been a fair share of studies justifying these correlations \citep{sev, cel}, both theoretical and empirical. Interestingly, the results seem to disagree wildly on the prominence of these correlations \citep{no_cor, no_cor2}, with some going as far as to questioning the signs of the correlations \citep{neg_cor}. \\

In this work, we define an object named {\it penal strategy} which factors into a person's utility function in the form of the expected level of punishment. Whenever an individual decides to commit a crime, he/she absorbs this disutility subconsciously into his/her utility function to evaluate whether committing crime is ``worth it" or not (see section \ref{gen}). Under a particular class of crime, a penal strategy can be specified by a vector of parameters $\bsph$. As an example, for the crime of DWI, a penal strategy can be specified with five parameters, $\bsph = \{p,f,t,\tau,r\} = \{10\%,\text{\$}200,\text{1 day},\text{6 hours},r\}$. Given this strategy, the offender is apprehended with probability $p=10\%$ (certainty of punishment), and he/she has the choice of being punished with a fine $f=\$200$ or be imprisoned for a duration of $\tau=\text{6 hours}$ (severity of punishment), with the delay in punishment being $t=\text{1 day}$ (celerity of punishment). \\

The parameter $r$ is an abstract quantity describing the rate at which disutility is incurred onto an individual during imprisonment (see section \ref{dis_sec}), and can be roughly interpreted as how ``unpleasant" the imprisonment condition is \citep{harsh}. One may be tempted to lump this quantity together with $\tau$ and define $r\tau$ simply as the total disutility incurred on the individual during imprisonment, thus effectively quantifying the ``severity" of punishment. Even ignoring the effect of hyperbolic time discount (see section \ref{dis_sec}), this coarse characterization of ``severity" makes sense only if the goal to focus on how severity affects the {\it personal} utility function of the offender. If we take into account the cost of {\it social} utility in actually implementing this strategy, $r$ is substantially different from $\tau$ in the sense that increasing $r$ (making the condition more unpleasant) incurs little cost while increasing $\tau$ (increasing the term of imprisonment) incurs a cost that scales roughly proportional to $\tau$ (see section \ref{soc_wel}). This seems to suggest that making the imprisonment condition more unpleasant is a more economically efficient strategy than increasing the term of imprisonment, which is true only to a certain extent, as we shall discuss in section \ref{solve_opt}\footnote{It has been suggested that prisons should be made as unpleasant as possible since making the prison more unpleasant (increasing $r$) costs essentially nothing \citep{harsh} while at the same time being able to decrease the level of criminal activity. However, this strategy is justified only if the well-being of the criminal is ignored as a factor for the social welfare function, then in this case the optimal strategy would trivially be a strategy that is infinitely unpleasant (e.g. execution). A more in-depth and practical study of the effect of $r$ on the total social benefit is presented in section \ref{sev_phase}}. \\

The discrepancy between the offender's personal utility function and the victim's utility function (even in the absence of the convolution of behavioral traits such as time discounting and risk preference) is the source of non-triviality for finding the optimal penal strategy (see section \ref{uti}). In fact, one can observe a rather nice {\it duality} between this discrepancy in the utility functions to the characterization of {\it specific} and {\it general} deterrence\footnote{There is a nice overview of these two modes of deterrence in the paper \citep{spe_gen}.}. {\it Specific deterrence} focuses on the effect of punishment targeting the offender {\it after} the crime has been committed. A strategy based on the principle of specific deterrence would then focus on the utility function of the offender, and the optimal penal strategy under this framework would be a strategy that incurs the minimal amount of disutility on the offender (while at the same time being sufficient enough to deter future offenses). On the other hand, {\it general deterrence} focuses on the effect of the {\it threat} of punishment on preventing any crimes from occurring in the first place. A strategy based on the principle of general deterrence would then focus on the {\it utility functions} of the victims and non-offenders, and the optimal penal strategy under this framework would be a strategy threatening a large amount of disutility on the offender. In an ideal world, this means that crime would occur very infrequently, which decreases the social cost of enforcing any punishment and increases the social benefit of the victims (see \ref{soc_wel} for a formal discussion). \\

It is then natural to let the function of which we try to find the maximum be the {\it social welfare function} \citep{sowel} which includes the sum of the utilities of all members in the population - the offenders, victims, and non-offenders. Under this definition of the optimization problem, the distinction between specific deterrence and general deterrence is non-existence from an utilitarian standpoint. As maximizing this function guarantees that the penal strategy is optimal in both frameworks of deterrence. The optimal penal strategy should then accomplishes three goals which all contribute to the maximization of the social welfare function, thus achieving an utilitarian optimum. The first goal is to maximize the total number of members that can be deterred from committing the crime in the first place; the second goal is to minimize the cost of implementing the strategy itself; and the third goal is to minimize the disutility incurred on the offender during the punishment. Note that this optimum, in general, does not correspond to the judicial optimum, where the level of punishment matches exactly with the degree of crime \citep{fair}, and the correlation between the two optima is an interesting point of research that we leave open for our future work.

\subsection{Hyperbolic Time Discounting}
\label{dis_sec}

Hyperbolic time discounting is a time-inconsistent model that accounts for the phenomenon that a delayed reward is generally less appealing to a human than an immediate reward, even though the reward is the same \citep{hyper}. There has been studies on how hyperbolic time discounting affects the behaviors of the offenders, and how it affects their perception of deterrence \citep{hyper1,hyper2}. However, the works are mostly empirically, and to our knowledge, there has been no work on how the phenomenon of hyperbolic time discounting can be actually used to inform an optimal penal strategy. \\

Formally, we can define a particular reward to be $A$, and the utility gain of receiving that reward delayed by $t$ to be $u(A,t)$. We can then express $u(A,t)$ as follow
\begin{equation*}
u(k,A,t) = \frac{u(A,0)}{1+kt},
\end{equation*}
where $u(A,0)$ is the utility gain of receiving the reward immediately. We see that $u(A,t)$ is a monotonously decreasing function with respect to $t$, with $k$ governing the rate of decrease. For the rest of the paper, we refer to $k$ as the {\it discount rate}\footnote{Technically speaking, this is an abuse of notation as the discount rate is only well-defined for the model of exponential time discounting, because the relative decay rate is constant for an exponential function but not for a hyperbolic function. Therefore, we use the terminology ``discount rate" purely for the sake of denoting the parameter $k$ in this context.}, which is a parameter that fully specifies the discounting function. The utility of a delayed reward can be easily generalized for modeling the disutility of a delayed punishment, where we simply interpret $u(A,t)$ as the disutility of a punishment, $A$, delayed by time $t$. \\

A more interesting application of the discounting function would be to evaluate the disutility of a continuous punishment, such as imprisonment, by integrating the discounting function with respect to time. To be more specific, we consider a continuous punishment where the disutility per unit of time is $-r$. If the punishment is enforced continuously in the time window $[t_1,t_2]$ on an offender whose discount rate is $k$, then the total disutility that the individual receives is
\begin{equation*}
u(k,r,t_1,t_2) = \int_{t_1}^{t_2}\, \frac{-rt}{1+kt}\,dt = -\frac{r}{k}\log\big[ \frac{1+kt_2}{1+kt_1} \big].
\end{equation*}
Alternatively, we can denote $t=t_1$ as the time duration from the current time to the start of the punishment, and $\tau = t_2-t_1$ as the length of the punishment. This allows us to write
\begin{equation}
\label{dis}
u(k,r,t,t+\tau) = -\frac{r}{k}\log\big[ 1+\frac{k \tau}{1+kt} \big].
\end{equation}
In the context of imprisonment, we can interpret $r$ as the disutility that the individual receives per unit time during imprisonment, including the opportunity cost of not working and the discomfort of the jail/prison environment. We denote $r$ as the {\it harshness} of the imprisonment condition. We can interpret $\tau$ as the term of imprisonment and $t$ as the time between getting caught to the punishment being enforced. In the language of deterrence theory, the pair $\{r,\tau\}$ quantifies the ``severity" of punishment, and $t$ quantifies the ``celerity" of punishment.\\ 

We can easily check the asymptotic behavior of $u(k)$ in the following two limits
\begin{equation*}
\lim_{k\to 0}u(k) = -r\tau \qquad \lim_{k\to\infty}u(k) = -\lim_{k\to\infty}\frac{r}{k}\log\big[ \frac{\tau}{t} \big] = 0.
\end{equation*}
For a person with a discount rate of $k=0$, the perceived disutility is simply the total disutiltiy of imprisonment in the absence of time discounting. As the discount rate $k$ increases, the magnitude of $u$ decreases monotonously, meaning that the threat of imprisonment is less effective for an individual with a higher discount rate. In the limit of large $k$, the disutility begins to scale inversely with the $k$, and approaches zero for ${k\to\infty}$. \\

Suppose we are implementing a strategy with the hope of incurring a disutility of $-u_0$ for the offender, then there the two parameters $\{t,\tau\}$ have to follow some constraint. To find this constraint, we set the expression in equation \ref{dis} equal to $-u_0$
\begin{equation}
\begin{split}
u(r,t,t+\tau) =& -\frac{r}{k}\log\big[ 1 + \frac{k\tau}{1+kt} \big] = -u_0 \\
\implies
\tau =& (e^{ku_0/r}-1)(\frac{1}{k} + t).
\end{split}
\end{equation}
Note that $\tau$ scales linearly with $t$, with the slope being $e^{ku_0/r}-1$, which scales exponentially with $k$. This means that when a punishment is delayed, the length of the punishment has to increase accordingly in order to ``compensate" for the disutility decay. For an individual with a higher discount rate $k$, the disutility decay as a result of the delay will be more prominent, so the length of punishment has to be increased by a greater amount to incur the desired disutility for the individual. In some sense, for an offender with a large discount rate $k$, the factor of ``celerity" is more important than ``severity". This discussion is formalized in appendix \ref{j2}.

\subsection{Probability Weighing Function}
\label{pros}

Having discussed the concept of hyperbolic time discounting and relating it to two of the three factors of deterrence - severity and celerity of punishment, we move on to the third factor of deterrence - certainty of punishment, which we should refer to more formally as {\it probability of apprehension}. This factor can be modeled under the economic framework of prospect theory \citep{prospect}. Simply put, prospect theory states that for an event with some known probability of occurring, the probability perceived by a human is different from the actual probability. Therefore, when an individual is presented with an uncertain reward, the expected utility gain that he/she perceives is not the actual expected utility, as exemplified by Allais paradox \citep{allais}. \\

Consider a simple lottery $L$ presented to an individual. The lottery returns reward $A$ with probability $p$, and returns reward $B$ with probability $(1-p)$, then the expected utility gain that the lottery returns is
\begin{equation*}
\E(u(L)) = pu(A) + (1-p)u(B).
\end{equation*}
This is, however, not the utility gain that the individual perceives. In fact, the perceived expected utility is given by
\begin{equation*}
U = \pi(p) u(A) + \pi(1-p) u(B),
\end{equation*}
where $\pi$ is commonly termed the {\it probability weighting function}, a function that attempts to model how human evaluates the utility of uncertain events. The probability weighting function can be expressed as
\begin{equation}
\label{pi}
\pi(p) = \frac{p^{\gamma}}{(p^{\gamma} + (1-p)^{\gamma})^{1/\gamma}},
\end{equation}
which is an inverse S shaped function with its level of curvature governed by the parameter $\gamma\in (0,1)$, which we term the {\it probability weighing factor}. The probability weighing factor appears to be different across different people and across different classes of rewards/punishments, but empirical estimates of its value generally fall within the range $[0.5,0.7]$ \citep{pros_review}. In section \ref{dist_prob}, we empirically estimate the mean of $\gamma$ among our survey sample to be $\overline{\gamma}\approx 0.61$ in the context of weighing the risk of committing DWI. Note that other forms of $\pi(p)$ do exist \citep{shape} which contains more than one parameters, but for this work, we focus on the form as appeared in equation \ref{pi}. \\

It can be easily checked that the function $\pi(p)$ satisfies the regularity condition of $\pi(0)=0$ and $\pi(1)=1$. This is required as impossible events should be perceived as impossible, and certain events should be perceived as certain. In addition to the regularity conditions, $\pi(p)$ is constructed such that there is another solution to $\pi(p)=p$ for $p\in (0,1)$. We can denote this solution as $p^*$, then it can be shown that (see appendix \ref{pi_prop})
\begin{equation*}
\begin{split}
\pi(p) > p \qquad \text{if } p<p^* \\
\pi(p) < p \qquad \text{if } p>p^*.
\end{split}
\end{equation*}
This corresponds to the phenomenon that humans tend to overweight low-probability events and underweight high-probability events. \\

Therefore, under a particular deterrent strategy with the probability of apprehension being $p$, the offender will perceive that probability as being $\pi(p)$. This discrepancy has to be taken into account when evaluating the probability of apprehension for the optimal penal strategy (see section \ref{final_opt}). To see this clearly, consider a punishment incurring a disutility of $-u_0$ on the offender, then the offender will perceive the expected disutility to be $-\pi(p)u_0$, instead of $-pu_0$. For $p<p^*$, the former is greater than the latter, meaning that the individual will over-evaluating the threat of punishment. The same argument goes for $p>p^*$, where the individual will under-evaluate the threat of punishment. From a very cursory cost-benefit analysis, this suggests that a deterrent strategy with a low probability of apprehension is more economically efficient. To see this, assume that the cost of apprehension is $c_pp$, which scales linearly with the probability of apprehension. We then compute the perceived disutility per unit of cost to be
\begin{equation*}
\label{pi_ratio}
\frac{\pi(p) u_0}{c_p p} \propto \frac{\pi(p)}{p},
\end{equation*}
which is a monotonously decreasing function for sufficiently small $p$ (see appendix \ref{pi_prop}). In fact, this ratio tends to infinity when $p\to 0$, which suggests a deterrence strategy with zero probability of apprehension. \\

This is obviously not right, and the reason why this simple analysis breaks down is because we failed to consider the full utility function of the offenders, which includes the utility gain returned from the crime itself (see section \ref{gen}). For now, equation \ref{pi_ratio} offers a nice intuitive insight as to why it may be preferable to use a deterrent strategy with low probability of apprehension, though it should be interpreted as a serious attempt for studying the cost-benefit ratio with respect to parameter $p$.

\subsection{Utility Function}
\label{uti}

There have been many studies investigating the correlations between criminal activity and obvious predictors such as wealth \citep{cor_w}, level of education \citep{cor_edu}, and social environment \citep{cor_soc}. Recently, there has been an emergence of studies offering behavioral analyses of the motivation behind criminal activity, which attempt to differentiate multiple facets of the behavioral traits including time-inconsistent behaviors \citep{cor_time}, risk averting/loving behaviors \citep{cor_risk}, and behaviors induced by underlying psychological factors \citep{cor_psy}. All these studies can be viewed as endeavors towards understanding the question of why certain individuals choose to commit crime under a particular deterrent strategy, while others do not. From an economic standpoint, the answer to this question is because the utility function of every individual is {\it parameterized} differently (see section \ref{gen}). In this work, we focus on mainly two important parameters $\{w,k\}$, wealth and discount rate. We believe that these two parameters are strong indicators of criminal behaviors (see section \ref{pop_part}), and the distribution of the two parameters can be accurately measured within a population (see section \ref{dist_dis} and \ref{dist_prob}). \\

An important side note to mention here is that it is, in fact, possible for $\{w,k\}$ to {\it fully} specify the behavioral traits of the population under two possible scenarios. The first scenario is when the pair $\{w,k\}$ forms a complete basis for describing all possible utility functions \citep{factor}, meaning that all other behavioral traits can be uniquely specified by the values of $\{w,k\}$. For example, it may be the fact the ``recklessness" of an individual \citep{reckless} is simply an expression of wealth and discount rate\footnote{Note that the direction of causality may go the other way, meaning that it may just as well be that the discount rate is just an expression of recklessness.}. An interpretation of this would be that if a person has a high discount rate, then he/she may be more ``reckless" in the sense that he/she will be prone to seek immediate rewards at the sacrifice of future well-being. This is generally an unrealistic scenario, as it is rather ridiculous to assume that the complexities of human behavior can be described by merely two parameters. \\

The second, more realistic scenario is to relax the assumption that $\{w,k\}$ has to fully specify the behavioral traits of {\it every} member in the population, but we instead only require it to describe the {\it average} behavioral traits of the population. Therefore, if other behavioral traits are distributed among the population independently with respect to $\{w,k\}$, then the general behavior of the population can be effectively predicted by $\{w,k\}$ ``on average". As an example, in section \ref{dist_prob}, we show empirically that the probability weighing factor $\gamma$, describing the risk preference of an individual, is in fact distributed independently with respect to $\{w,k\}$. In the end, if we wish to find the optimal penal strategy, we are only interested in how the criminal behaviors across the entire population will respond to a particular deterrent strategy (see section \ref{pop_part}), so it is only required that we know how the behavioral traits are distributed among the population (see section \ref{dist_tr}), meaning that it is unnecessary to predict the response of every single member.

\subsubsection{Utility Function of the Offender}

We first begin by assuming that all individuals are {\it rational agents}\footnote{The question of whether the assumption that all individuals are rational agents is valid or not is somewhat meaningless in a mathematical sense, as all ``irrational" behavior of an individual can be explained, or made ``rational" essentially, by a particular parameterization of the utility function such as his/her discount rate, probability weighing, underlying psychological factors, etc.}. According to rational choice theory, whenever an individual is faced with a decision among multiple choices, the individual will select the one returning the highest utility gain (or lowest utility loss) for him/herself. This theory also applies to the decision process of a potential offender who weighs the expected return of the crime to decide whether to commit the crime or not. In short, the individual will only choose to commit the crime if the {\it perceived} expected utility gain is greater than zero. We can denote $\bs{\theta}$ as the vector of parameters defining the utility function of the individual (which we simply refer to as {\it utility parameters} from here on), and denote $\bs{\phi}$ as the current penal strategy. We let the utility gain returned from the crime be $B(\bs{\theta})$, and the perceived expected disutility of the possible punishment be $-D(\bs{\theta},\bs{\phi})$. The individual will be deterred from committing the crime if the net utility gain of the crime is negative, or
\begin{equation*}
\label{cur_det}
B(\bsta) - D(\bsta,\bsph) < 0.
\end{equation*}
Note that the utility parameters $\bsta$ are different for every individual, so given a particular deterrent strategy $\bsph$, the above inequality is satisfied for certain values of $\bsta$ but violated for others. Therefore, equation \ref{cur_det} (at equality) can be interpreted as a partition that separates the population into offenders and non-offenders (see section \ref{pop_part} for a detailed discussion for this partition). \\

As a simple example, we consider the case where the parameter $\bsta$ is fully specified by the discount rate $k$. Imagine a scenario where two individuals, Alice and Bob, are given the opportunity to commit a crime with a guaranteed reward returning some fixed utility gain $b_0$. Alice has a small discount rate $k_A$, and Bob has a large discount rate $k_B > k_A$. If the crime is committed, there is some fixed change of being apprehended with the punishment being a 24 hour imprisonment starting next week. According to equation \ref{dis}, we see that $D(k_A)>D(k_B)$ (as $k_A < k_B$), which means that Alice perceives the disutility of the punishment to be greater than what Bob perceives. Assuming that Alice is deterred from committing the crime, then
\begin{equation*}
b_0 - D(k_A) < 0 \centernot\implies b_0 - D(k_B) < 0,
\end{equation*}
meaning that Bob won't necessarily be deterred as well. Intuitively, what this means is that a punishment delayed by a week for Bob may not present sufficient level of threat for Bob than it does for Alice, so the deterrent strategy is not an effective strategy for Bob. 

\subsubsection{Utility Function of the Victim}

A very crucial assumption here is that the offender is {\it selfish} in the sense that he/she seeks only to maximize his/her utility function and disregards the utility loss for the victim. If we denote the disutility of the victim as a result of the crime to be $-L(\bsta)$ (where $\bsta$ is the utility parameter for the victim), then we define {\it total utility function} to be the sum of the utility functions of both parties of the crime (the offender plus the victim)
\begin{equation*}
B_{tot}(\bsta_1,\bsta_2) = B(\bsta_1) - L(\bsta_2),
\end{equation*}
where $\bsta_1$ and $\bsta_2$ are the utility parameters for the offender and victim respectively. Note that the offender will act to maximize only $B$ but not $B_{tot}$. Therefore, a naive deterrent strategy would be to set the expected disutility of punishment for the offender to be exactly equal to the disutility of the victim as a result of the crime, or $L(\bsta_2) = D(\bsta_1,\bs{\phi})$, which translates to the problem of solving for $\bsph$. Under this deterrent strategy, the offender implicitly evaluates the disutility incurred on the victim, and would act indirectly to increase the total utility function\footnote{Note that under this model, a crime is justified as long as the net change in $B_{tot}$ is positive after the offense. A commonly invoked example is when a poor offender steals from a rich victim, forcing a more efficient distribution of wealth, as the marginal utility of a dollar for the offender is greater than that of the victim.}. \\

This is a rather elegant solution, but it fails in mainly two respects. First, the solution to the equation $L(\bsta_2) = D(\bsta_1,\bsph)$ depends on the parameters $\bsta_1$ and $\bsta_2$. In other words, a different deterrent strategy has to be implemented for each offender-victim pair, which is neither practical nor fair, as it implies using some behavioral traits of the offender (which may be poorly measured) as a discriminant for enforcing a particular punishment. Second, even under the assumption that this strategy is possible and the total utility function $B_{tot}$ between every offender-victim pair is maximized, the implementation of this strategy may be very costly, with its cost outweighing the net increase in $B_{tot}$. A general extension of this analysis is presented in section \ref{gen}, with the analysis being performed under a general social welfare function which encapsulates the total utility functions of all the offender-victim pairs as well as the cost of implementing the penal strategy itself.

\section{General Model of Deterrence}
\label{con_opt}

At this point, it should be very clear that the problem of finding an optimal strategy is not a trivial task, as it requires a simultaneous maximization the total utility function of each offender-victim pair and minimization of the cost of implementing the strategy. As briefly discussed in section \ref{det_sec}, the problem can be defined as finding a strategy that maximizes the {\it social welfare function}, which we define to be the sum of the total utility of every member in a population. Note that the absolute social welfare function is very difficult to define and measure \citep{wel_meas}, so we focus only on the relative change in the social welfare function connected to the particular crime we are studying\footnote{This includes any victim's loss associated with the crime, the cost that went into deterring this crime, and the disutility resulting from the punishment of the offenders of this crime.}, which we refer to as the {\it relative social welfare function}. In this paper, the crime we are focusing on is the offense of driving under influence, or DWI or short, then the goal is to find a penal strategy $\bsph = \{f,t,\tau,p,r\}$ (see section \ref{det_sec}) such that it maximizes the relative social welfare function of the crime of DWI.

\subsection{Review of Old Model}

The notion of formulating the problem of finding an optimal deterrent strategy in terms of maximizing the social welfare function is not novel. A systematic formulation has been developed by Gary Becker in his seminal paper which employs economic methods to determine the optimal probability and severity of punishment \citep{becker}. However, there are multiple problems that the original model failed to address, and we here point out the three major ones. \\

First, the model assumes that the cost of enforcing the punishment scales proportionally with the disutility of the punishment that the offender perceives. This is generally not the case. For example, if we double the term of imprisonment, the cost of enforcing the imprisonment may be doubled, but the disutility for the offender may not be due to the presence of hyperbolic time discounting (see equation \ref{dis}). Second, the model assumes that the punishment is either a fine or imprisonment, and the offender does not have a choice between the two. This assumption is rather problematic as it suggests the ideal strategy being a fine of a very large sum of money, which would ideally deter all offenders while costing very little (as the collection cost of fines is assumed to be small). In reality, a large amount of fine is rather meaningless for individuals of low income, as they would not be able to afford it anyway. A more realistic model will be to offer the offender the {\it option} to choose between a fine, imprisonment, and other forms of punishments. Lastly, the model accounts for the fact that an increase in the probability of apprehension and severity of punishment will reduce the number of offenses. However, it does not offer a formulation that allows for estimating exactly how many offenses would be deterred. Therefore, the optimization problem formulated only exists on a theoretical level, and a closed form solution containing variables that can actually be measured in a population does not exist. 

\subsection{Overview of New Model}
\label{gen}

In our model, we characterize the behavioral traits of each member with the utility parameters $\bsta$ (see section \ref{uti}), and we attempt to describe the criminal propensity of a population by the distribution of $\bsta$ among a population (see section \ref{dist_tr}), which allows us to predict the response of the population to a particular penal strategy. To be more specific, our model is based on the assumption that a strategy will partition the population into offenders and non-offenders (see section \ref{pop_part}), and the social welfare function can be explicitly evaluated over this partition, so the problem is converted in some sense to the optimal partitioning of the population into offenders and non-offenders. In addition, the model also assumes that the offender has the freedom to choose between multiple punishments, and for each possible punishment, both the social cost of enforcement and the disutility incurred on the offender is carefully considered (see section \ref{soc_wel}). \\

We first present a rigorous mathematical formulation of the relative social welfare function that can be constructed with respect to any class of crimes. The idea is to assume every ordered pair of two members within a population to be a possible pair of offender and victim. And under a specific deterrent strategy, we can divide all the pairings into two main groups. The first group consists of pairings on which a criminal offense won't be realized due to successful deterrence, and this group generates social benefits from the crimes not occurring, which factor into the social welfare function as positive summation terms. The second group consists of pairings on which the deterrence fails and the occurrence of criminal offenses is possible, and this group generate social costs associated with the enforcement of punishments on the offenders, which factor into the social welfare function as negative summation terms. The model is general but often times quite difficult to study, but in most cases, simplifications can be made under reasonable assumptions justified under the crime of interest (see section \ref{ass}). 

\subsection{Opportunities and Offenses}

To begin with, we define a {\it population} $\Omega$ to be a set \citep{set} whose elements $\omega\in\Omega$ are members making up the population. We define $\mu$ to be the counting measure operator \citep{measure} which returns the cardinality of a finite set, and we denote the size of the population to be $\mu(\Omega) = |\Omega| = N$. For every member $\omega_i \in \Omega$, we can fully characterize the member with its utility parameters $\bsta_i$. \\

We define a particular class of {\it crime} to be $\zeta$. Given a crime $\zeta$, and some time window $\Delta t$, a directed graph \citep{graph} whose vertices are $\Omega$ is generated, with the arrows representing a potential offender-victim pair. More specifically, the arrow $(\omega_i,\omega_j)$ represents an {\it opportunity} for crime $\zeta$ with $\omega_i$ being the {\it offender} and $\omega_j$ being the {\it victim}. We can model the arrival of opportunities at each edge to be an independent Poisson process \citep{probability} with rate $2\lambda$, with the direction of the arrow being random. Therefore, the expected number of opportunities presented to every member $\omega_i\in \Omega$ in the time window $\Delta t$ is approximately $N \lambda \Delta t$\footnote{Since the arrival of opportunities at each edge is a Poisson process with rate $2\lambda$, the formation of an arrow pointing from $\omega_i$ to some other vertex is also a Poisson process with rate $\lambda$. Therefore, the formation of an arrow pointing from $\omega_i$ to any of all other $N-1$ vertices is a sum of $N-1$ independent Poisson processes, with the expected value being $(N-1)\lambda\Delta t \approx N\lambda\Delta t$ for large $N$.}. \\

We denote a {\it punishment} to be $\delta$, which is defined by the form\footnote{The form of the punishment can be a simple fine, imprisonment, community service, etc.}, the celerity, and the severity of the punishment. We define the set of all possible punishments be $\Delta$. In addition, we define the probability of apprehension to be $p\in[0,1]$. We can then express the penal strategy as $\bsph = (p,\delta_1,\delta_2,...,\delta_m)$, which is an $(m+1)$-tuple. Under this strategy, the offender will be apprehended with probability of $p$ and given $m$ choices of punishment as specified by $\delta_1,\delta_2,...,\delta_m$. The set of all deterrent strategies is then given by the Cartesian product $\bs{\Phi} = [0,1]\times \Delta^{m}$\footnote{There are overlapping elements in the set $\bs{\Phi}$, as the penal strategy is not affected by the ordering of $(\delta_1,...,\delta_m)$. This means that there are $m!$ elements in the set corresponding to the same strategy.}. 

\subsection{Deterrence}

Note that when opportunity arrives for an offeder $\omega_i$, the offender won't necessarily take the opportunity. The opportunity will only be realized and converted into an {\it offense} if the net utility gain $U_{\zeta}$ from committing the crime is positive for the potential offender under the deterrent strategy $\bsph$. We can then write the condition for deterrence as
\begin{equation}
\label{det}
U_{\zeta}(\bsta_i,\bsta_j,\bsph) = B_{\zeta}(\bsta_i,\bsta_j) - \pi(p,\gamma_i)\big\{ \min_{\delta\in\bsph}\big[ D_{\zeta}(\bsta_i,\delta) + S_{\zeta}(\bsta_i,\delta) \big] \big\} < 0,
\end{equation}
where $B$ denotes the utility that the offender $\omega_i$ expects to gain from the crime $\zeta$, which is dependent on both $\bsta_i$ and $\bsta_j$ (the traits of the offender and the victim). $\pi(p,\gamma_i)$ is the probability weighing function of the offender parameterized by $\gamma_i$. $D+S$ denotes the total disutility that the offender expects to be incurred on him/her after apprehension, where $D$ denotes the disutility of punishment $\delta$ for the offender $\omega_i$\footnote{Note that $D_{\zeta}$ is only present in the utility function of the offender if he/she is actually aware of the deterrent strategy. See section \ref{unin} for a detailed discussion of the case where the offender is uninformed.}; and $S$ denotes any disutility of being apprehended not directly resulting from the punishment (such as the stigma associated with being apprehended). For the rest of the paper, we refer to $S$ simply as {\it stigma} \citep{stigma}. Note that the individual will always choose the punishment incurring the least disutility, hence the $\min$ function. \\

On the other hand, we can write the net gain/loss in the social welfare function as a result of the offense to be
\begin{equation*}
V_{\zeta}(\bsta_i,\bsta_j) = B_{\zeta}(\bsta_i,\bsta_j) - L_{\zeta}(\bsta_i,\bsta_j)
\end{equation*}
where $-L(\bsta_i,\bsta_j)$ denotes the disutility incurred on the victim from the offense. Note that the utility that the offender gains from the crime\footnote{It is assumed that the utility that the offender actually gains from the crime is equal to the utility that he/she expects to gain when evaluating his/her utility function before deciding to commit the crime} is taken into account as part of the social welfare function (since the social welfare function must include the utility function of the offender as well). This means the increase in the social welfare function as a result of a crime not occurring is simply the inverse of $V$, or
\begin{equation}
\label{soc}
-V_{\zeta}(\bsta_i,\bsta_j) = L_{\zeta}(\bsta_i,\bsta_j) - B_{\zeta}(\bsta_i,\bsta_j).
\end{equation}

\subsection{Costs of Punishment}
\label{cost_pun}

To model the social cost of implementing the deterrent strategy, we consider two separate costs. The first cost is associated with the apprehension probability, which we can model as a fixed cost that depends only on the apprehension probability, $C_p(p)$, which we refer to as the {\it detection cost}. And the second cost is associated with enforcing the punishment on the offender, which depends on the form of punishment and also the two parties involved, $C_{\delta}(\bsta_i,\bsta_j,\delta)$. For the sake of simplicity, we do not consider any reformation effects as a result of the punishment (see section \ref{ass} for cases where this is approximately true). In other words, we assume that the offender's utility parameters will not be modified through deterrence, so there is no social benefit associated with the decrease in likelihood of recidivism \citep{recid}. To study the social cost associated with the punishment more closely, we consider the punishments of fine and imprisonment as examples. For the sake of simplicity, we assume that the social cost of punishment depends only on the utility parameters of the offender. 

\subsubsection{Costs of Fine}

We first consider the example of a simple fine, where the only cost associated with this punishment is the collection cost\footnote{We are making the general assumption that the transfer of wealth from the offender to the collector is a simple transfer of utility within the population, so it does not result in any utility gain/loss in the social welfare function. This means that we are ignoring the effect of any inefficient distribution of wealth \citep{dist_w} that may result from this trasnfer, and the only cost associated with this transfer is the cost of forcing this transfer itself.} and the disutility is due to the offender's stigma associated with apprehension. We can then write the total social cost of using a fine as punishment as
\begin{equation}
\label{fine_cost}
C_{\delta}(\bsta_i,f) = C_f + g S_{\zeta}(\bsta_i,f),
\end{equation}
where $f$ denotes the punishment of a fine, and $C_f$ denotes the collection cost assumed to be independent of the fine amount. We denote $C_f$ to be the {\it enforcement cost} corresponding to the cost of enforcing the punishment onto the offender. $g>1$ is a prefactor translating the offender's stigma into the disutility incurred on the society. To see the reasoning behind this prefactor and why it is greater than one, we can assume that the offender is a productive worker, and this stigma prevents the offender from being fully productive in the future\footnote{This is due to the offender not being able to find a suitable job due to his/her criminal records.}. We set the loss in the social utility (due to the deviation of the offender's output from his/her full capability) as $g S$, then we see that this value must be greater than the loss for the offender in his/her potential compensation for his/her work, $S$\footnote{This is because in a realistic economic model, the worker is only compensated with a fraction of his output value, with the rest of the value being absorbed by the consumer and the company.}. We can denote $g S$ as the {\it opportunity cost} of the punishment, corresponding to the value (unrelated to the crime) that the offender is otherwise able to create if he/she were not apprehended. 

\subsubsection{Costs of Imprisonment}

We then consider the social cost of imprisonment. Note that besides the cost of having the stigma associated with apprehension, we also have to consider social costs such as the opportunity cost of keeping the offender in prison (instead of letting him/her work) for a time period of $\tau$, and also the social benefit of the offender not being able to commit any more crime for the duration of his/her term of imprisonment. We can then write the total social cost of using imprisonment as punishment as
\begin{equation}
\label{imp_cost}
C_{\delta}(\bsta_i,I) = C_I(t,\tau) + g\big[ D_{\zeta}(\bsta_i,I) + S_{\zeta}(\bsta_i,I) \big] + \lambda\tau \sum_{\omega_j} V_{\zeta}(\bsta_i,\bsta_j),
\end{equation}
where $I$ denotes the punishment of imprisonment, and $C_I$ is the enforcement cost which can be assumed to be a function of $t$ (the delay in punishment) and $\tau$ (the length of imprisonment), with an explicit expression of this cost given in equation \ref{ci}. In addition to the opportunity cost $gS$ incurred {\it after} the period of punishment, $gD$ can be interpreted as the opportunity cost incurred {\it during} the punishment as a reuslt of the offender not working for a time duration and $\tau$ plus any ``discomfort" he perceives from the unpleasant imprisonment environment\footnote{If we assume the presence of time discounting, then the social disutility as a result of imprisoning the offender ($D$ in equation \ref{imp_cost}) will be substantially different than the disutility that the offender perceives ($D$ in equation \ref{det}). To be more specific, the $D$ in equation \ref{imp_cost} will not be time discounted as it represents actual social disutility, while the $D$ in equation \ref{det} will be time discounted as it represents the offender's perception of punishment. This discrepancy is very important if we were to use the discount rate $k$ as a parameter of $\bsta$ when partitioning the population and evaluating the social cost of imprisonment (see sections \ref{pop_part} and \ref{soc_wel}).}. The reasoning for the prefactor $g$ is similar to the discussion in the previous paragraph. And the last term denotes the social benefit of the criminal opportunities not being realized as a crime for a time period of $\tau$. 

\subsection{Population Partition}

We first define the following set of all ordered pairs $E = \{ (\omega_i,\omega_j) \,|\, \omega_i\in\Omega,\, \omega_j\in\Omega,\, \omega_i\neq\omega_j \}$, which corresponds to all possible offender-victim pairs. We may partition this set into multiple subsets given a deterrent strategy $\bsph = \{p,\delta_1,...,\delta_m\}$. We denote the first subset as $E_0 \subseteq E$, which satisfies the following condition
\begin{equation*}
\forall (\omega_i,\omega_j)\in E_0, \quad U_{\zeta}(\bsta_i,\bsta_j,\bsph) < 0.
\end{equation*}
In other words, when an opportunity is formed from $\omega_i$ to $\omega_j$, the offense will not be realized as the net utility gain for the offender is negative. The complement of this subset $E\setminus E_0$ denotes the set of pairs on which offenses will be realized, and it can be further partitioned into $m$ disjoint subsets, with $m$ being the number of punishment options. For every $h\in[[1,m]]$, the subset $E_h \subseteq E$ satisfies the following condition
\begin{equation*}
\forall (\omega_i,\omega_j)\in E_h, \quad U_{\zeta}(\bsta_i,\bsta_j,\bsph) \geq 0 \,\land\, \text{argmin}_{\delta\in \bsph}\big[ D(\bsta_i,d) \big]=\delta_h.
\end{equation*}
In other words, whenever an opportunity is formed from $\omega_i$ to $\omega_j$, an offense will be realized as the net utility gain for the offender is positive. Furthermore, when caught, the offender will choose $\delta_h$ as punishment. 

\subsection{Social Welfare Function}
\label{gen_wel}

We are now finally in the position to construct the social welfare function particular to the crime $\zeta$. We first zero the social welfare function to the welfare level corresponding to the scenario where every criminal opportunity has been realized, and no deterrent strategy has been implemented, then the problem of maximizing the expected social welfare function with respect to all possible deterrent strategies $\bsph$ is given by (where the subscript $\zeta$ is assumed)
\begin{equation}
\label{opt_og}
\max_{\bsph\in \bs{\Phi}} \Big\{ \lambda\Delta t\Big[ \sum_{(\omega_i,\omega_j)\in E_0} \big( -V(\bsta_i,\bsta_j) \big) - p\sum_{h=1}^m \sum_{(\omega_i,\omega_j)\in E_h} C_{\delta}(\bsta_i,\bsta_j,\delta_h) \Big] - C_p(p) \Big\},
\end{equation}
Note that $\bsph$ enters implicitly in the expression through the partitioning of the set $E$ and the cost associated with the punishment of offenders $C_{\delta}$. \\

In this form, the model is complete. However, this formulation is not useful in any practical sense due to two reasons. First, in order for the optimization problem to be defined, we have to specify the distribution of $\bsta$ in the population $\Omega$, otherwise there would be no way to ``count" the number of members in each of the $m+1$ partitions of $E$. The second problem is that the optimization problem is not computationally feasible. If we analyze the time complexity \citep{complex} of performing this optimization brute force, the algorithm has to check all possible deterrent strategies, with the number of operations scaling exponentially with $m+1$\footnote{This is assuming some sort of discretization of the set of all possible deterrent strategies $\bs{\Phi}$}. Furthermore, given a penal strategy, the partitioning of $E$ requires $N^2$ operations\footnote{This is because all edges have be to be checked, and the number of edges for a complete graph of $N$ vertices is $N(N-1) = O(N^2)$.}. Therefore, the time complexity of a brute force algorithm would be $O(N^2 e^{m+1})$, which presents a major computational challenge. 

\section{A DWI Case Study}
\label{dwi_case}

As discussed in section \ref{gen_wel}, the general form of the optimization problem as appeared in equation \ref{opt_og} is intractable from a computational standpoint. However, if we assume that the optimization problem is performed in the context of a specific crime, many simplifying assumptions can usually be made. In this section, we focus on the crime of DWI as an illustrative case study.

\subsection{Simplifying Assumptions}
\label{ass}

If we let the crime $\zeta$ be the class of DWI crimes, then there are several assumptions we can make to simplify the optimization problem as given in equation \ref{opt_og}, to the point where an analytic approach is possible. There are two things to note. First, these assumptions are not strictly necessary for the main results of the paper to hold. Second, alternative simplifications can also be made depending on the crime of interest. We here make assumptions that are justified particular to the crime of DWI. \\

We first state the seven assumptions that we make, followed by a detailed discussion of the justification of these assumptions
\begin{itemize}
\item{\it {\bf Assumption One:} Every member can be fully characterized by his/her level of wealth and discount rate.}
\item{\it {\bf Assumption Two:} The utility function of every member is static.}
\item{\it {\bf Assumption Three:} The probability weighing factor is constant for the population.}
\item{\it {\bf Assumption Four:} The utility gain from an offense for the offender is dependent on his/her level of wealth, and the utility loss from an offense for the victim is fixed in expected value. }
\item{\it {\bf Assumption Five:} The penal strategy is static and non-discriminatory is expected value. }
\item{\it {\bf Assumption Six:} The opportunity cost of imprisonment, $S+D$, scales proportionally with the offender's level of wealth. In addition, the opportunity cost incurred during the punishment, $D$, scales proportionally with the term of imprisonment $\tau$. }
\item{\it {\bf Assumption Seven:} The stigma associated with apprehension is independent of the form of punishment. }
\end{itemize}

{\bf Assumption one} means that $\bsta = \{w,k\}$ can be fully specified with two parameters, where $w$ denotes the level of wealth and $k$ denotes the discount rate. We believe that these two parameters describe well the criminal behaviors of a population (see section \ref{dis_sec}), and can be easily measured within a population (see section \ref{emp}). For an individual, the level of wealth correlates strongly with the utility gain from committing DWI (to be discussed shortly) and the opportunity cost of imprisonment (see section \ref{cost_pun}). The discount rate governs how fast the disutility of punishment decays with its delay (see section \ref{dis_sec}). In some sense, this pair of parameters allows us to determine the ``level" of deterrence that the member perceives from a penal strategy, allowing us to predict his/her response when given a criminal opportunity. Furthermore, the disutility incurred on the member as a result of the punishment can also be easily evaluated from the two parameters. \\

{\bf Assumption two} means that the utility function of the individual does not evolve in time, or at least in the time window in which the penal strategy is in effect. In the context of DWI, this means that as long as the punishment of multiple offenses is equal to the first offense, the offender won't be less inclined to commit DWI even after an accident or apprehension. In other words, recidivism is possible for the offender. Even if we were to relax this assumption and allow the utility function of an offender to be modified after an accident or apprehension, if we assume that the probability of accident or apprehension is sufficiently small or the time interval between two consecutive DWI offenses is at the same order as the penal strategy time window\footnote{If the probability of accident or apprehension is small, then we are only mismodeling a small fraction of the population whose utility functions are modified after the incident. Similarly, if the rate of DWI offenses is small, then we can assume that the rate of change in the utility function is also small, allowing us to make the static assumption \citep{adi} in the time frame where the penal strategy is implemented.}, then the change in the utility function will only account for a second order correction to the partitioning of the population, which we can ignore. \\

{\bf Assumption three} states that the probability weighing factor $\gamma$ is the same for every individual. Although this is not necessarily a realistic assumption (as seen in section \ref{dist_prob}), it is crucial for the simplification of the optimization problem as the function $\pi(p,\gamma)$ is related to $\gamma$ in a highly nonlinear fashion (see equation \ref{pi}). \\

{\bf Assumption four} is a reasonable assumption in the context of drunk driving. Whenever an individual decides to commit DWI, in the majority of cases, it is usually motivated by the fact that the alternative of not committing DWI will result in some disutility scaling positively with his/her level of wealth\footnote{A typical example would be when a person decides to drive home from a bar after drinking, the corresponding alternative would be to call a taxi and pick up his/her vehicle the next day, and the time spent doing so will result in the decrease of total productivity correlated positively with his/her current level of wealth} \citep{dwi}. Furthermore, it can be assumed that probability and severity of the damages resulting from the DWI act is independent of $\bsta$. To begin with, the purpose of performing DWI (unlike other crimes such as burglary) is not for the forceful transfer of wealth from the victim to the offender, so there is no reason to assume that the expected disutility incurred on the victim will depend on the utility parameters of either party\footnote{The involvement in a car accident is generally unintentional, with the probability and severity of damage correlated with the ability of the driver to operate the vehicle under influence, which can be reasonably assumed to be independent of $\bsta$. The damage can be set fixed in expected value, which is simply the expected damage of the car accident weighted by the probability.}. This allows us to set the expected loss to the victim to some fixed amount. \\

{\bf Assumption five} means that the probability of apprehension and punishment choices are the same for every offender regardless of his/her $\bsta$. However, DWI incidents may result in varying level of damages, ranging from minor damages such as running into a pole to serious damages such as killing a pedestrian. The punishment obviously has to be discriminatory towards the level of damages to be in accordance with criminal justice, meaning that offenders whose DWI act resulted in severe damages should be punished more severely. Nevertheless, as discussed in the last paragraph, the expected damage is fixed for every $\bsta$, so we can also make the assumption that the expected punishment will also be the same every $\bsta$. This implies that the enforcement cost of the punishment should be constant in expected value. \\

{\bf Assumption six} states that the total loss in the value of the offender's labor output (as a result of imprisonment) scales proportionally with his/her level of wealth. This is a fair assumption as a member's level of wealth should act as a strong indicator of his/her rate of productivity. And since the total value of output is simply the rate of productivity multiplied by the time period, it can also be assumed that the opportunity cost incurred during the imprisonment period scales proportionally with the term of imprisonment. \\

{\bf Assumption seven} implies that the negative effects of having a DWI record itself is the same regardless of punishment options that the offender chooses. This makes sense because the severity of the crime itself should be the only substantial characterization of the criminal background of the offender, and the form of punishment (or even the severity of punishment) should play little role in affecting the future prospect of the offender. \\

Under these assumptions, equations \ref{det} and \ref{soc} simplify to
\begin{equation}
\label{soc_red}
\begin{split}
U_{\zeta}(\bsta_i,\bsph) &= B_{\zeta}(\bsta_i) - \pi(p) \big\{ \min_{\delta\in \bsph}\big[ D_{\zeta}(\bsta_i,\delta) \big] + S_{\zeta}(\bsta_i) \big\} < 0 \\
V_{\zeta}(\bsta_i) &= B_{\zeta}(\bsta_i) - L_{\zeta}
\end{split}
\end{equation}
respectively, noting that the functions are independent of $\bsta_j$, or the victim's utility parameters. And this allows for considerable simplification for the social welfare function in equation \ref{opt_og}. \\

For further simplification, we can assume that a fine and imprisonment are the two only punishment options $\Delta = \{f,I\}$, or $\delta_1 = f$ and $\delta_2=I$. The the social welfare function can be reduced to
\begin{equation}
\label{opt_new}
\begin{split}
& N\lambda\Delta t\Big[ \sum_{(\omega_i,\omega_j)\in E_0} \big(-V(\bsta_i,\bsta_j)\big) - p\sum_{h=1}^m \sum_{(\omega_i,\omega_j)\in E_h} C_{\delta}(\bsta_i,\bsta_j,\delta_h) \Big] - C_p(p) \\
\sim& \sum_{\omega_i\in\Omega_0} \big(-V(\bsta_i)\big) - 
p \sum_{w_i\in\Omega_1} C_{\delta}(\bsta_i,f)
- p \sum_{w_i\in\Omega_2} C_{\delta} (\bsta_i,I)
 - \frac{C_p(p)}{N\lambda\Delta t} \quad \text{(Divide by $N\lambda\Delta t$)} \\
=& \sum_{\omega_i\in\Omega_0} \big[ L-B(\bsta_i) \big] 
- p\sum_{\omega_i\in\Omega_1}\big[ c_f + gS(\bsta_i) \big]  \\
-& p\sum_{\omega_i\in\Omega_2}\Big\{ C_I(t,\tau) + g\big[ S(\bsta_i)+D(w_i,\tau) \big] 
- \lambda\tau N\big[ L-B(\bsta_i) \big] \Big\} - \frac{C_p(p)}{N\lambda\Delta t}
\end{split}
\end{equation}
where the domain of optimization $\bsph\in\bs{\Phi}$ and the subscript $\zeta$ is assumed. To understand this reduction,  First, we first realize that neither the summands nor the partitions depend on $\omega_j$ (the victim) anymore, so the summation over $\omega_j$ is factored out to give us a prefactor of
\begin{equation*}
\sum_{\omega_j\neq \omega_i} = \mu(\Omega)-1 \approx N.
\end{equation*}
Since the partition depends only the offender $\omega_i$, we can define the partition over the vertices $\Omega$ instead of the pairs $E$\footnote{The partitioning is defined through the first equation in \ref{soc_red}, which is only dependent on $\omega_i$.}. We then denote the three subsets as $\{\Omega_0,\Omega_1,\Omega_2\}$\footnote{$\Omega_0$ corresponds to non-offenders. $\Omega_1$ corresponds to offenders choosing a fine over imprisonment. $\Omega_2$ corresponds to offenders choosing imprisonment over a fine.}, with the number of subsets being $3=1+2$ as there are only two choices of punishments (fine or imprisonment). Recall that this optimization problem is over five parameters $\bsph = \{p,f,t,\tau,r\}$ (see section \ref{det_sec}), which is still somewhat computationally expensive. We show in the following sections how we can reparameterize the penal strategy to better specify the partitioning of the population (section \ref{pop_part}), and how certain assumptions can be assumed to reduce the optimization to its asymptotic form (section \ref{repar}).
 
\subsection{Population Partition}
\label{pop_part}

There are several factors that enter the mindset of a potential offender performing DWI. The first is simply the utility gain from the crime itself, which we can assume to scale proportionally with his/her level of wealth (see \ref{ass})
\begin{equation*}
B(\bsta) = bw,
\end{equation*}
where $b$ is denoted as the scaling factor. Recall that the probability of apprehension is $p$, which the offender evaluates to $\pi(p)$, and the offender has the choice between a fine or imprisonment. For a fine, we can denote the disutility to be $f$, and for imprisonment, we can write its disutility magnitude as (see equation \ref{dis})
\begin{equation}
\label{imp_dis}
I(w,k,t,\tau) = \frac{rw}{k}\log\big[ 1 + \frac{k\tau}{1+kt} \big],
\end{equation}
where the expression is scaled with $w$. Recall that $\tau$ is the length of imprisonment, and $t$ is the time of the delay in punishment. \\

If we model the stigma as $-sw$, then the member will be deterred from committing a DWI offense if the following is satisfied (see equation \ref{det})
\begin{equation}
\label{dwi_det}
U(w,k) = bw - \pi(p) \Big\{ \min\big[ f, I(w,k,t,\tau) \big] + sw \Big\} < 0,
\end{equation}
where we've applied the assumption that the incurred stigma is independent of the form of punishment (hence why the min function is only over $f$ and $I$). The above inequality can be expressed equivalently as
\begin{equation}
\label{wk}
w < \frac{\pi(p)f}{b-\pi(p)s} \quad \land \quad
\frac{k}{\log\big[ 1+\frac{k\tau}{1+kt}\big]} < \frac{\pi(p) r}{b-\pi(p)s}.
\end{equation}
If we denote 
\begin{equation}
\label{w0}
w_0 = \frac{\pi(p)f}{b-\pi(p)s},
\end{equation}
then the first condition can be written as $w < w_0$. And since $k/\log\big[ 1+\frac{k\tau}{1+kt} \big]$ is a monotonously increasing function with respect to $k$, there must be only one solution to the second condition in equation \ref{wk} at equality, which we can denote as $k_0$\footnote{Note that no analytic expression of $k_0$ exists at $k_0$ is the solution to a transcendental equation.}. Then similarly, we are allowed to write the second condition as $k < k_0$. \\

We then see that in order for an individual to be deterred, its level wealth must be below $w_0$ {\it and} its discount rate must be below $k_0$. We then say that the penal strategy is {\it targeting} a wealth level of $w_0$ and a discount rate of $k_0$. This defines the first subset of population $\Omega_0$ (see section \ref{gen}), which we term the {\it non-offenders}. We can interpret the subset $\Omega_0$ as consisting members whose level of wealth is sufficiently low to be deterred by a fine of amount $f$, while simultaneously having a sufficiently small discount rate to be deterred by an imprisonment length of $\tau$ (delayed by $t$). Note that given a strategy targeting $w_0$, the amount of fine is specified by 
\begin{equation*}
f = \frac{w_0\big[ b-\pi(p)s \big]}{\pi(p)}.
\end{equation*}
And given a strategy targeting $k_0$, the pair $\{t,\tau\}$ must be related as follow
\begin{equation}
\label{taut}
\tau = (e^{\psi k_0}-1)(\frac{1}{k_0}+t),
\end{equation}
where the $\psi$ is defined as follow
\begin{equation}
\label{psi}
\psi = \frac{b-\pi(p)s}{\pi(p)r}.
\end{equation}

An important side note to mention is that in a realistic population sample, there should always be some positive infimum to the set of wealth levels which we can denote as $w_m$, or
\begin{equation*}
\inf\{ w \,|\, w\in\Omega \} = w_m > 0.
\end{equation*}
This assumption is necessary to account for some minimal wage standard of the society, and to make it mathematically possible to model the wealth distribution as a Pareto distribution \citep{pareto}, for which there must be a positive lower bound to the domain of the distribution (see section \ref{dist_tr}). We then consider a possible strategy where the targeted wealth level is smaller than the minimum wealth level, or $w_0<w_m$. In this case, every member would be an offender (or the non-offender subset, $\Omega_0$, would be empty), as the disutility of the fine is too small to deter a member of even the lowest level of wealth. For the sake of optimization, we don't have to consider this strategy as this strategy only incurs enforcement cost, but does not return any social benefit as none of the members are deterred by the strategy. \\

\begin{figure}
\centering
\includegraphics[scale = 0.8]{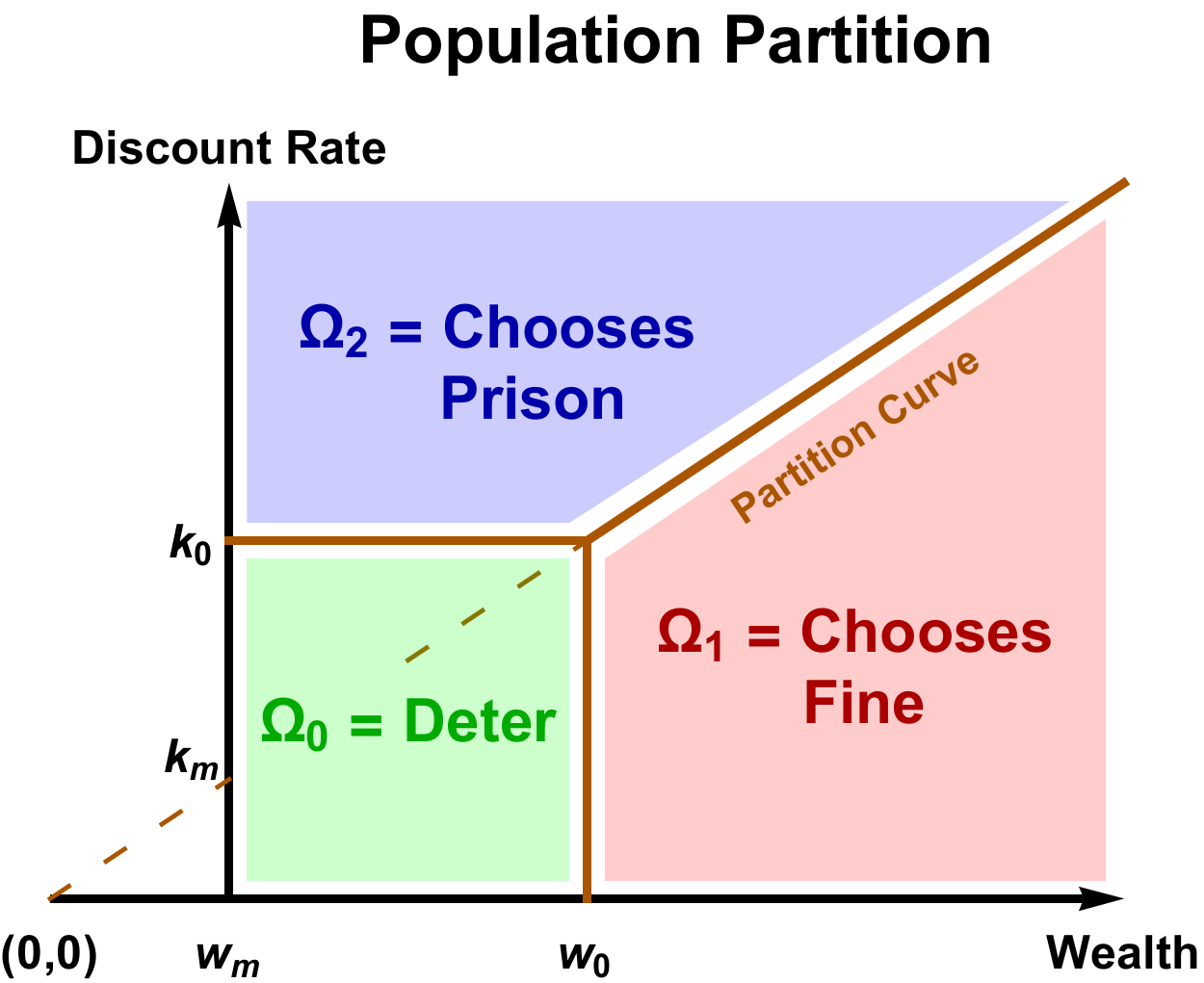}
\caption{\label{plot_part} A visual representation of the partitioning of population under a penal strategy targeting $\{w_0,k_0\}$. The green rectangle, on the lower left, represents the population deterred from committing any crime. The red trapezoid, on the right, represents the offenders that will choose a fine over imprisonment. The blue trapezoid, on the top, represents the offenders that will choose imprisonment over a fine. Color in print.}
\end{figure}

The {\it offenders} are members in the population for which the penal strategy fails, and the set of offenders can be denoted as the complement of the set of non-offender, or $\Omega/\Omega_0$. The set of offenders can be further partitioned into two subsets depending on whether the offender chooses a fine or imprisonment after being apprehended. The partition can be expressed as a curve on the $w-k$ coordinate system on which the members are indifferent towards the two punishment options. This is done by simply taking the two arguments of the $\min$ function in equation \ref{dwi_det} to be equal, which gives us
\begin{equation}
\label{curve}
f = I(w,k,t,\tau) 
\implies w = \frac{kf}{r\log\big[ 1+\frac{k\tau}{1+kt} \big]}.
\end{equation}
The above equation defines a curve in the $w-k$ coordinate system dividing the population into two regions. On one side of the curve, the population prefers a fine over imprisonment; we denote this subset as $\Omega_1$. On the other side of the curve, the population prefers imprisonment over a fine; we denote this subset as $\Omega_2$. We call this curve the {\it partition curve}. See figure \ref{plot_part} for a visual representation of the partition. Note that the offenders choosing a fine over imprisonment have a relatively low discount rate and high level of wealth, and the offenders choosing imprisonment over a fine have a relatively high discount rate and low level of wealth. \\

In most cases, we can approximate the partition curve as a straight line
\begin{equation}
\label{part_app}
\frac{w}{k} \approx \frac{w_0}{k_0},
\end{equation}
noting that this expression does not depend explicitly on $\{t,\tau\}$. See appendix \ref{asy} for a discussion on how the approximation is carried out. The partitioning scheme is then completely defined by the pair $\{w_0,k_0\}$ (see figure \ref{plot_part}), which provides a huge simplification from the original 5 degrees of freedom of the penal strategy $\{f,t,\tau,p,r\}$. From equation \ref{taut}, we see that given $k_0$ and $t$, $\tau$ can be uniquely specified, so the deterrent strategy can be alternatively parameterized as $\bsph = \{p,w_0,k_0,t,r\}$. The purpose of using $\{w_0,k_0\}$ as parameters is to reduce the complexity of the optimization problem by allowing the boundaries of the partitions to be easily described by the two parameters (see section \ref{dist_tr}).

\subsubsection{Uninformed Members}
\label{unin}

The majority of the analysis performed in this section can be easily extended to model the population that is {\it uninformed} of the law \citep{unin}. If a member is uninformed, then he/she is unaware of the deterrent strategy, thus the total disutility $D_{\zeta}$ will be absent from his/her utility function, which gives us
\begin{equation*}
U(w,k) = (b-\pi(p)s)w.
\end{equation*}
We can make the simplifying assumption that $b>s$ (meaning that the utility gain from the crime is always greater than the stigma associated with apprehension), then the utility function is strictly positive. Therefore, the uninformed member will always choose to commit the crime given the opportunity. \\

This means that the uninformed members won't contribute to the subset $\Omega_0$, since all of them are assumed to be offenders. However, when apprehended, an uninformed member would be given the same choices of punishments (as an informed member would be), so the partitioning between $\Omega_1$ and $\Omega_2$ still applies to the uninformed population exactly as it would for the informed population.

\subsection{Social Welfare Function}
\label{soc_wel}

Recall from section \ref{gen} that given a particular strategy $\bsph$, the social welfare function can be written as the social benefit gained from the deterred crimes under the strategy minus the detection cost and social cost of punishment (which is further broken down into enforcement cost and opportunity cost). \\

We first focus on the social benefit of deterred crimes. Whenever an individual commits an offense, a loss is usually incurred on the society. Using equation \ref{soc_red}, we expressed the expected utility gain/loss of the society after an offense to be
\begin{equation*}
V(w) = bw - l
\end{equation*}
where $bw$ is the increase in utility of the offender, $-l$ is the expected disutility incurred on the victim (assumed constant). If we zero the utility function to be when all criminal opportunities are realized and no penal strategy is in effect (see section \ref{gen}), then the utility gain/loss of the society after an offense being deterred is simply the inverse of $V$, or
\begin{equation*}
-V(w) = l - bw.
\end{equation*}

We now consider the costs associated with implementing the strategy. We first look at the detection cost, which we can assume to scale proportionally with the probability of detection
\begin{equation*}
C_p(p) = c_p p,
\end{equation*}
where $c_p$ is the scaling factor\footnote{A more realistic cost model for the probability of apprehension would be $C_p(p) = c_p\log\big[ \frac{1}{1-p} \big]$ (Note that this expression diverges for $p=1$, corresponding to the fact that certain apprehension is not possible in reality.) The derivation of this cost model is based on the fact that the failure of apprehension should decay exponentially with the number of ``checkpoints", which should scale proportionally with the cost. However, for $p \ll 1$, this can be approximated with $C_p(p)\approx c_p p$.}. \\

Lastly, we consider the social costs of the punishments. From equation \ref{fine_cost}, we see that the social cost of using a fine as punishment can be written as
\begin{equation*}
c_f + gws,
\end{equation*}
where $c_f$ is the fixed collection cost and $gsw$ is the opportunity cost. Similarly, from equation \ref{imp_cost}, we write the social cost of using imprisonment as punishment as
\begin{equation*}
C_i(t,\tau) + gw(s+r\tau) - \lambda\tau N(l-bw),
\end{equation*}
where $gw(s+r)$ is the opportunity cost of imprisonment, and $\lambda\tau n(l-bw)$ is the social benefit of the offender not being able to commit crimes for a time duration of $\tau$. For the enforcement cost $C_i(t,\tau)$ of imprisonment, we require that
\begin{equation*}
\partial_t C_i < 0 \quad \partial_{\tau}C_i > 0,
\end{equation*}
which implies that the cost increases with duration of imprisonment and decreases with the length of delay\footnote{The reason why we assume that the enforcement cost decreases with the length of delay is an empirical assumption. Consider a scenario where the punishment of imprisonment is enforced with a delay rather than being immediate, then the party enforcing the punishment will have sufficient time to better allocate the resources necessary for the enforcement, so that the enforcement will be performed in a more economically efficient fashion. In addition, there should also be the legal cost associated with sentencing the offender, and we can assume that the cost to increase with the rate at which the legal process is carried out.}. A possible cost model of $C_i(t,\tau)$ that satisfies the two conditions above can be written as
\begin{equation}
\label{ci}
C_i(t, \tau) = c_0 + \frac{c_t}{mt} + c_{\tau}\tau
\end{equation}
where $c_0$ is some fixed cost of enforcing the punishment that does not depend on $t$ or $\tau$. $\frac{c_t}{mt}$ can be interpreted as the ``celerity" cost of the punishment that scales inversely with $t$. The choice of using $\frac{c_t}{mt}$ as the ``celerity" cost is for analytic convenience and is not entirely realistic, though the exact choice of the celerity cost model does not alter the major results of this work\footnote{In this work, we are only interested in the asymptotic scaling behavior of the social costs of imprisonment (see appendix \ref{optcost}) with respect to $k_0$, which should be at least exponential regardless of celerity cost model we use (as long as $\partial_t C_i<0$). To see this, simply consider a cost model without celerity cost, or $C_i(t,\tau)=c_0+c_{\tau}\tau$, which should give us the best scaling with respect to $k_0$ (as it does not incur any additional cost to decrease the delay in punishment). We simply set $t=0$, then from equation \ref{taut}, we see that $\tau=(e^{\psi k_0}-1)/k_0$ scales exponentially with $k_0$, which implies that $C_i$ must also scale exponentially. Therefore, for any celerity cost model satisfying $\partial_{\tau}C_i < 0$, the social costs of imprisonment must scale {\it at least} exponentially with $k_0$ as well.}. And $c_{\tau}\tau$ is the ``severity" cost which scales proportionally with the duration of imprisonment $\tau$. \\

Putting everything together, we see that the total social costs associated with each punishment option (fine and imprisonment) can be written as
\begin{equation}
\label{cd}
C_{\delta}(w,t,\tau,\delta) = 
\begin{cases}
c_f + gsw \quad &\text{if } \delta = \delta_1 \\
c_0 + \frac{c_t}{mt} + c_{\tau}\tau + gw(s+r\tau) - \lambda\tau N(l-bw) \quad &\text{if } \delta = \delta_2,
\end{cases}
\end{equation}
where $\delta_1$ is associated with the punishment of a fine, and $\delta_2$ is associated with the punishment of imprisonment. Therefore, equation \ref{opt_new} can be expressed explicitly as
\begin{equation}
\label{opt_dwi}
\begin{split}
&\sum_{\Omega_0}(l-bw) \\
-& p\sum_{\Omega_1}(c_f + gsw) \\
-& p\sum_{\Omega_2}\big[ c_0 + \frac{c_t}{mt} + gsw + (c_{\tau} - \Lambda l)\tau + (gr + \Lambda b)w\tau \big] \\
-& c_p p
\end{split}
\end{equation}
where we've denoted $\Lambda = N\lambda$. Note that the prefactor $N\lambda\Delta t$ is ignored and simply absorbed into the constant $c_p$\footnote{This gives us $c_p=\frac{c_p}{N\lambda\Delta t}$, which can be interpreted as the cost of apprehension per unit probability, per unit time, per unit population, per unit criminal activity.}. At this point, we have accomplished two major task. The bounds of the summation indices (or the population partition in the $w-k$ space) are uniquely specified by $\{w_0,k_0\}$ (see \ref{pop_part}), and the summands are expressed in terms of $\{p,t,\tau,r\}$ (with $t$ and $\tau$ constrained as equation \ref{taut}). However, to perform the summation, we still have to know exactly how $\{w,k\}$ is distributed among each subset $\{\Omega_0,\Omega_1,\Omega_2\}$. In other words, we have to define the {\it measure} over which the summation is performed \citep{measure}. This is done in the immediately following section, by assuming a distribution of $\{w,k\}$ over the population.

\subsection{Modeling the Distribution of Traits}
\label{dist_tr}

The optimization can be made simpler if we approximate the counting measure as a probability measure scaled by the total population $N(1+\epsilon)$, where $\epsilon$ accounts for the uninformed population\footnote{It is assumed that the distribution of $\{w,k\}$ over the informed and uninformed population will be the same.}. In other words, the number of members with traits $\{w,k\}$ can be approximated as
\begin{equation*}
N \times f_{\{W,K\}}(w,k)\,dw\,dk,
\end{equation*}
where $f(w,k)$ can be interpreted as the ``density" of population with traits $\{w,k\}$. If we further assume that $w$ and $k$ are distributed independently (with an empirical justification presented in section \ref{inde}), then $f(w,k)$ can be factorized as
\begin{equation*}
f_{\{W,K\}}(w,k) = f_W(w)f_K(k).
\end{equation*}
We can model the distribution of wealth, $f_W(w)$, as a Pareto distribution \citep{pareto} with the domain being $w\in(w_m,+\infty)$, where $w_m$ is the minimum wealth level of the population
\begin{equation}
\label{pdf1}
f_W(w) = \frac{\alpha w_m^{\alpha}}{w^{\alpha+1}} \qquad F_W(w) = 1-(\frac{w_m}{w})^{\alpha}.
\end{equation}
In addition, we can model the distribution of discount rate, $f_K(k)$, as a zero-inflated\footnote{Note that the use of the term ``zero-inflated" is not technically valid, as the model is a continuous distribution instead of a discrete one. The term is rather used to denote the fact a large probability measure is concentrated at $k=0$.} exponential distribution \citep{inflate}
\begin{equation}
\label{pdf2}
f_K(k) = (1-\rho) \delta(k)+\rho\frac{1}{\beta}\exp(-\frac{k}{\beta}) \qquad
F_K(k) = (1-\rho) + \rho\big[ 1-\exp(-\frac{k}{\beta}) \big],
\end{equation}
where $\delta$ is the Dirac Delta function\footnote{The Dirac Delta function $\delta(k)$ is a function that evaluates to zero everywhere except at $k=0$, at which it is undefined. However, the Dirac Delta function has a well-defined integral which evaluates to one, or $\int_{-\infty}^{\infty}\delta(k)\,dk=1$. Alternatively, $\delta(k)$ can be interpreted as a distribution that generates a probability measure that is completely localized at $k=0$. The use of $\delta(k)$ is for the convenience of modeling the high concentration of probability measure at $k=0$. Its use is not strictly necessary. In fact, any distribution that generates a high concentration of probability measure close to $k=0$ would suffice} \citep{dirac}, which accounts for the phenomenon that the majority of the population is expected to have a discount rate close to zero (meaning that most people are ``rational" in time). The parameter $\rho$ is the fraction of population with non-zero discount rate, and $\beta$ is the mean discount rate among that population. The form of this distribution is justified with empirical measurements in a sample population of size 207 in section \ref{dist_dis}, where it is shown that the discount rates of the sample is well fitted by this distribution, and the estimators for $\{\rho,\beta\}$ from the sample are given. \\

Having now modeled the distribution of $\{w,k\}$, we now give the explicit form of the social welfare function in integration form. For the sake of convenience, we first denote the following ratios
\begin{equation}
\label{ratio}
u = \frac{k_0}{w_0} = \frac{k_m}{w_m} 
\qquad
v = \frac{k_m}{k_0} = \frac{w_m}{w_0}.
\end{equation}
Under the assumption that $w_0>w_m>0$, or equivalently $v\in(0,1)$, the social welfare function can be approximated as follow
\begin{equation}
\label{opt}
\begin{split}
\Omega_0 \qquad \rightarrow \qquad &\int_{w_m}^{w_0}\int_0^{k_0}\,(l-bw)\,f(w,k)\,dk\,dw \\
\Omega_1 \qquad \rightarrow \qquad &- p\,\Big\{ \int_{w_0}^{\infty} \int_0^{uw}\,(c_f+gsw)\,f(w,k)\,dk\,dw + \epsilon\int_{w_m}^{\infty}\int_0^{uw}\,(c_f+gsw)\,f(w,k)\,dk\,dw \Big\} \\
\Omega_2 \qquad \rightarrow \qquad &- p\,\Big\{ \int_{k_0}^{\infty} \int_{w_m}^{k/u}\,C_{\delta}(w,t,\tau,\delta_2)\,f(w,k)\,dw\,dk + \epsilon\int_{k_m}^{\infty}\int_{w_m}^{k/u}\,C_{\delta}(w,t,\tau,\delta_2)\,f(w,k)\,dw\,dk \Big\} \\
&- c_p p,
\end{split}
\end{equation}
where again, we are ignoring the prefactor of $N$ by absorbing it into the constant $c_p$ (see section \ref{soc_wel}) and the expression for the social costs of imprisonment, $C_{\delta}(w,t,\tau,\delta_2)$ is given in equation \ref{cd}. A visual representation of the bounds of the integrals is given in figure \ref{plot_part}. Note that at this point, the optimization problem is completely well defined over the parameters $\{p,w_0,k_0,t,r\}$, and the following section will be devoted to solving this optimization through a combination of the techniques of further reparameterization and asymptotic analysis.

\section{Solving the Full Optimization Problem}
\label{solve_opt}

In this section, we perform an explicit optimization on the social welfare function. Even though the tools used for performing the optimization contains certain mathematical technicalities, there is a central concept that can be readily understood on an intuitive level. In short, we show that the pair $\{f,r\}$ (the fine amount and the harshness of imprisonment condition) is able to together induce an interesting {\it phase transition} on the optimal penal strategy. Section \ref{phase} provide an informal discussion of the concept of {\it phase transition} which originates from statistical physics \citep{phase}. In short, we show that when $r$ is above a certain threshold (or when the imprisonment condition is sufficiently harsh) and $f$ is below a certain threshold (or when the amount of fine is sufficiently small), then the social welfare function favors a penal strategy with a very severe term of imprisonment; otherwise, a severe term of imprisonment is not favored.

\subsection{Phase Transition}
\label{phase}

The term {\it phase transition} \citep{phase} is mainly used in physics to commonly describe the transition between different states of matter such as from solid to liquid (melting) or from liquid to gas (boiling). We here make no attempt to formally define the concept of phase transition as it is not completely relevant to our main points of discussion. In fact, we use the term {\it phase transition} very loosely in the context of penology to describe the phenomenon where when certain penal parameters cross a threshold, the space of {\it realizable}\footnote{A realizable penal strategy means one that can be implemented without an excessively large amount of social cost} penal strategies experiences a sudden change. This is analogous to the natural phenomenon where as a thermodynamic parameter (such as temperature) crosses a threshold \citep{therm}, the state of the matter changes discontinuously (such as water boiling at 100 degrees). Here, our ``thermodynamic parameter" are $\{f,r\}$, and when it crosses a certain threshold, the value $k_0$ at the optimal penal strategy rises abruptly to an extremely high value, corresponding to the phenomenon where a severe strategy suddenly becomes favorable.

\subsection{Reparameterization of the Social Welfare Function}
\label{repar}

Note that in section \ref{dist_tr}, we assumed that the factorization of the pdf of $\{w,k\}$ is possible, or $f_{\{W,K\}}(w,k) = f_W(w)f_K(k)$. This means that the integrals over $f_{\{W,K\}}(w,k)$ (see equation \ref{opt}) can also be factorized as separate integrals over $w$ and $k$, which allows for considerable simplification. If we denote the integral corresponding to subset $\Omega_i$ as $J_i$ (meaning that the social welfare function can be written as $J_0 - pJ_1 - pJ_2 - c_pp$), then we find the following closed form expressions for the three integrals (where the denotations in equation \ref{ratio} are applied)
\begin{equation*}
\begin{split}
J_0 &= (1-\rho e^{-\kappa_0})\Big[ l(1-v^{\alpha}) - \frac{\alpha}{\alpha-1}bw_m(1-v^{\alpha-1}) \Big] \\
\\
J_1 &= c_f\Big\{ v^{\alpha}\big[ 1-\alpha\rho E_{\alpha+1}(\kappa_0) \big] + \epsilon\big[ 1-\alpha\rho E_{\alpha+1}(\kappa_m) \big] \Big\} \\
&+ gs\Big\{ \frac{\alpha}{\alpha-1}(v^{\alpha}w_0 + \epsilon w_m) - \alpha\rho\big[ v^{\alpha}w_0 E_{\alpha}(\kappa_0) + \epsilon w_m E_{\alpha}(\kappa_m) \big] \Big\} \\
\\
J_2 &= \rho\big[ c_0 + \frac{c_t}{mt} + (c_{\tau} - \Lambda l)\tau \big]\,
\Big\{ e^{-\kappa_0}  - v^{\alpha}\kappa_0 \, E_{\alpha}(\kappa_0) +\epsilon\Big[ e^{-\kappa_m} - \kappa_m \, E_{\alpha}(\kappa_m) \Big] \Big\} \\
&+ \rho\frac{\alpha}{\alpha-1}\big[ gs+(gr+\Lambda b)\tau \big] w_m \,
\Big\{ e^{-\kappa_0} + \epsilon\, e^{-v\kappa_0} - (v\kappa_0)^{\alpha-1}\big[ \Gamma(2-\alpha, \kappa_0) + \epsilon\, \Gamma(2-\alpha,v\kappa_0) \big] \Big\}
\end{split}
\end{equation*}
where the following denotations are made
\begin{equation*}
\kappa_0 = \frac{k_0}{\beta} \qquad \kappa_m = \frac{k_m}{\beta}.
\end{equation*}
Note that $E_{\alpha}(x)$ denotes an exponential integral and $\Gamma(2-\alpha,x)$ denotes the upper Gamma function with their respective integrals (an asymptotic approximation) given by
\begin{equation*}
\begin{split}
E_{\alpha}(x) &= \int_1^{\infty} \frac{e^{-tx}\,dt}{t^{\alpha}} \approx 
e^{-x}(x^{-1} - \alpha x^{-2}) \\
\Gamma(\alpha,x) &= \int_x^{\infty} t^{\alpha-1}e^{-t}\,dt \approx
x^{\alpha-1}e^{-x}\big[ 1+(\alpha-1)x^{-1} \big]
\end{split}
\end{equation*}
The integrals can then be further reduced by applying these asymptotic approximations
\begin{equation*}
\begin{split}
J_0 &= (1-\rho e^{-\kappa_0})\Big[ l(1-v^{\alpha}) - \frac{\alpha}{\alpha-1}bw_m(1-v^{\alpha-1}) \Big] \\
\\
J_1 &\approx c_f\,\Big\{ v^{\alpha}\big[ 1-\alpha\rho \frac{e^{-\kappa_0}}{\kappa_0}\big] + \epsilon \big[ 1-\alpha\rho \frac{e^{-v\kappa_0}}{v\kappa_0} \big] \Big\} \\
&+ gs\,\Big\{ v^{\alpha-1}w_m\big[ \frac{\alpha}{\alpha-1} - \alpha\rho\frac{e^{-\kappa_0}}{\kappa_0} \big] + \epsilon w_m\big[ \frac{\alpha}{\alpha-1} - \alpha\rho\frac{e^{-v\kappa_0}}{v\kappa_0} \big] \Big\} \\
\\
J_2 &\approx \rho\big[ c_0 + \frac{c_t}{mt} + (c_{\tau}-\Lambda l)\tau \big]\,
\Big\{ (1-v^{\alpha})e^{-\kappa_0}+\alpha v^{\alpha}\frac{e^{-\kappa_0}}{\kappa_0} + \alpha\epsilon \frac{e^{-v\kappa_0}}{v\kappa_0} \Big\} \\
&+ \rho\frac{\alpha}{\alpha-1}\big[ gs + (gr+\Lambda b)\tau \big] w_m \,
\Big\{ \big[ (1-v^{\alpha-1})e^{-\kappa_0} + (\alpha-1)v^{\alpha-1}\frac{e^{-\kappa_0}}{\kappa_0} + (\alpha-1)\epsilon\frac{e^{-v\kappa_0}}{v\kappa_0} \big]\Big\}.
\end{split}
\end{equation*}
If we look at the optimization problem above, we immediately note that the pair $\{t,\tau\}$ appears only in the expression of $J_2$. In other words, they only contribute to the social cost of enforcing imprisonment as the punishment. This means that the optimization over $\{t,\tau\}$ can be done independently for $J_2$ to reduce one degree of freedom as a preliminary for the full optimization to be performed. \\

In appendix \ref{optcost}, we discuss in detail how this optimization is performed, and show that for a given penal strategy targeting $\kappa_0$, the optimal delay in punishment $t$ decays exponentially with $\kappa_0$ and the length of imprisonment $\tau$ scales exponentially with $\kappa_0$. This means that both the celerity and severity of punishment has to increase exponentially with the target discount rate $\kappa_0$, which agrees with our intuition. The minimum social cost of imprisonment $J_2$ decays exponentially with $\kappa_0$, with the asymptotic rate of decay being
\begin{equation}
\label{decay}
\frac{\psi\beta}{2}-v < 0,
\end{equation}
where the expression of $\psi$ is given in equation \ref{psi}. \\

Note that if the condition \ref{decay} is satisfied, then the social cost of imprisonment $J_2$ decays to $0$ for $\kappa_0\to\infty$, and the social cost of a fine $J_1$ converges to some fixed value. If we further assume that \begin{equation*}
\frac{c_t}{m}\big[ c_{\tau} - \Lambda l + (gr+\Lambda b)w_m \big] \gg c_f^2,
\end{equation*}
which essentially means that the social costs of imprisonment is much greater than that of a fine, then it can be easily seen that the maximum of the social welfare function can be approximated by taking $\kappa_0\to\infty$, corresponding to a very severe term of imprisonment. On the other hand, if the condition \ref{decay} is not satisfied, then $J_2$ diverges for $\kappa_0\to\infty$, so the maximum of the social welfare function is at some intermediate value of $\kappa_0$. In other words, the condition \ref{decay} can be interpreted as the threshold between a severe and mild term of imprisonment. \\

Note that in most practical scenarios, condition \ref{decay} can be easily satisfied (see sections \ref{sev_phase} and \ref{harsh}). For the sake of simplicity, we then assume in this case where \ref{decay} is satisfied and take $\kappa_0\to\infty$, which reduces the social welfare function to
\begin{equation}
\label{opt_reduce}
\begin{split}
& J_0 - p(J_1+J_2) - c_p p \\
=& l(1-v^{\alpha}) - \frac{\alpha}{\alpha-1}bw_m (1-v^{\alpha-1}) - pc_f (v^{\alpha} + \epsilon) - \frac{\alpha}{\alpha-1}pgsw_m( v^{\alpha-1} + \epsilon ) - c_p p.
\end{split}
\end{equation}
Note that are left with only two parameters $\{v,p\}$, comparing this to the original five $\bsph = \{p,f,t,\tau,r\}$. This is a much simpler optimization problem and can be approached analytically. But before we proceed with solving this problem, we first have a closer look at the condition \ref{decay} and see what it implies for the penal strategy.

\subsection{Threshold for Phase Transition}
\label{sev_phase}

Note that by plugging in the expression for $\psi$ in equation \ref{psi}, condition \ref{decay} becomes
\begin{equation}
\label{pq1}
\begin{split}
v >& \beta\frac{b-\pi(p)s}{2\pi(p)r}.
\end{split}
\end{equation}
In addition to the above condition, $v$ and $p$ also have to be within the real interval $[0,1]$\footnote{Note that $p$ denotes a probability, so $p\in [0,1]$ by definition. Recall that $v=\frac{w_m}{w_0}$ in equation \ref{ratio}, and $w_0>w_m>0$, implying that $0<v<1$.}. Therefore, we see that in order for condition \ref{pq1} to be satisfiable for $v,p\in [0,1]$, the inequality must hold for $\{v,p\} = \{1,1\}$, which implies that
\begin{equation}
\label{r_cond}
1 > \beta \frac{b-s}{2r}
\implies
\boxed{r > \frac{(b-s)\beta}{2}}.
\end{equation}
This means that the harshness $r$ must be at least $\frac{(b-s)\beta}{2}$, where we recall that $\beta$ is the mean discount rate of the population, and $b-s$ is the difference between the utility gain of the offense and the stigma of apprehension (per unit wealth). We see that the harshness must scale proportionally with $b-s$. To interpret this, note that if $b-s$ is positive, it means that the stigma of apprehension alone is not sufficient to deter the offender from committing DWI. Furthermore, the larger this difference is, the more ``prone" to DWI the offender would be, so the harshness of the imprisonment condition $r$ must increase accordingly to ``match" this difference, such that its presence in the offender's utility function is prominent enough to achieve the deterrent effect. In addition, note that $\beta$ is the mean discount rate of the population. For a population with a large $\beta$ value, we can interpret this as the population being generally ``reckless", so naturally, we must also increase the harshness of the imprisonment condition to match the level of recklessness of the population. \\

\begin{figure}
\includegraphics[scale=0.85]{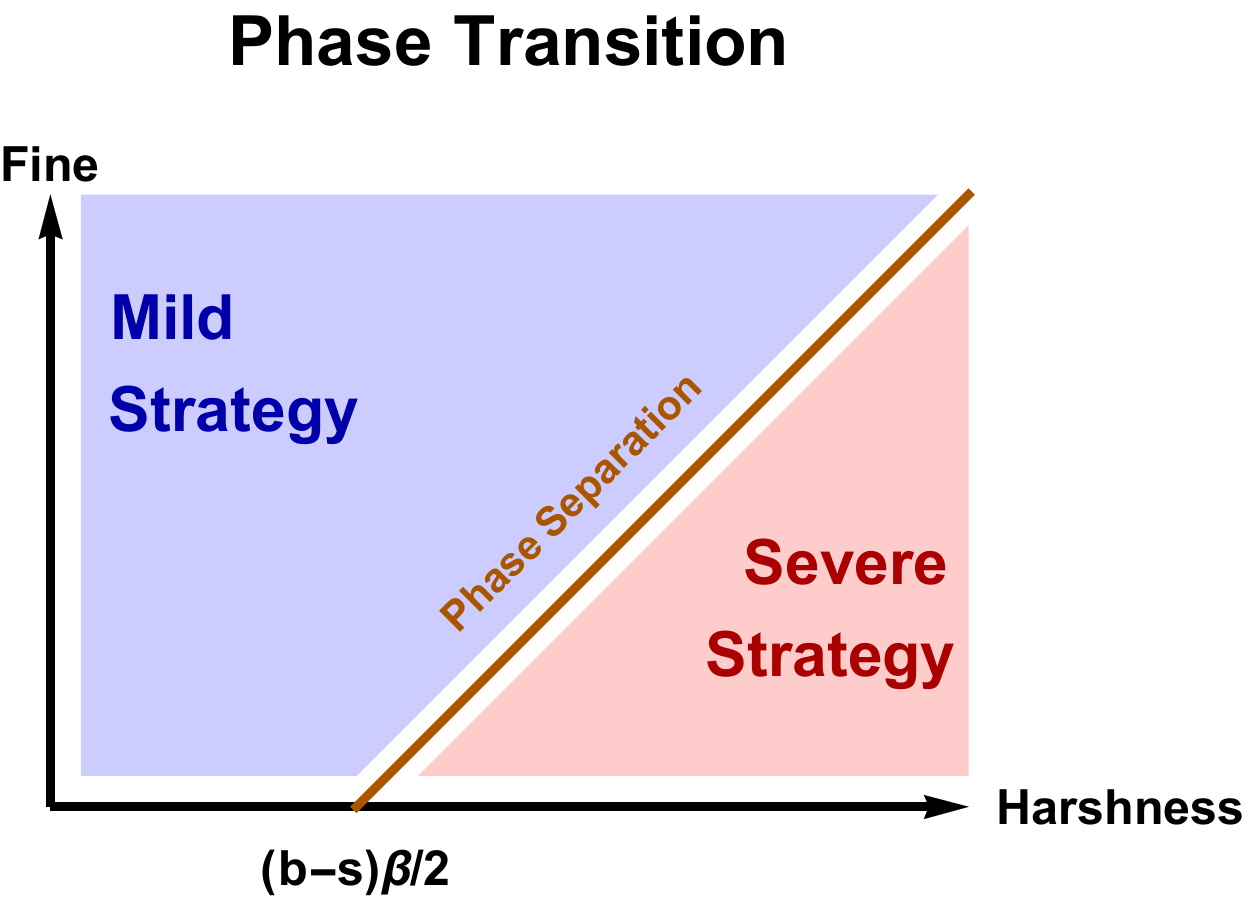}
\centering
\caption{\label{plotphase} Note that the optimal strategy experiences a very abrupt transition from a strategy having a small $k_0$ value (a mild term of imprisonment) to a strategy having a high $k_0$ value (a severe term of imprisonment). The transition occurs when the harshness of the imprisonment condition increases above a certain threshold $r>\frac{(b-s)\beta}{2}$, and the amount of fine decreases below a certain threshold $f<\frac{2rw_m}{\beta}$. Color in print.}
\end{figure}

To express condition \ref{pq1} in the original parameters of the penal strategy, we plug the ratio $v=\frac{w_m}{w_0}$ and the expression for $w_0$ (equation \ref{w0}) into inequality \ref{decay}, which gives us
\begin{equation}
\label{f_cond}
\boxed{
f < \frac{2rw_m}{\beta},
}
\end{equation}
which implies an upper bound on the amount of fine. Recall that $w_m$ denotes the minimum level of wealth among the population, and $\beta$ denotes the mean discount rate among the population with non-zero discount rates. Note that
\begin{equation}
I(w_m,\beta,0,(2e-1)\beta^{-1}) = \frac{rw_m}{\beta}\log \big[ 1 + \beta\beta^{-1}(2e-1) \big] = \frac{2rw_m}{\beta},
\end{equation}
which is the disutility of an immediate imprisonment of length $(2e-1)\beta^{-1}$ incurred on a member of $\{w_m,\beta\}$ (minimum level of wealth, and mean discount rate). This means that $f$ has to be small enough such that when the option of a fine of amount $f$ and an immediately imprisonment of length $(2e-1)\beta^{-1}$ is presented to this member, the member will choose a fine over imprisonment. Although the derivation of this bound is rather non-trivial, the intuition behind the need to limit the fine amount is clear. If the amount of fine is too high, then the majority of the population will opt for imprisonment instead. And if there is an overabundance of offenders choosing imprisonment over a fine, and this makes it unfavorable to implement a strategy with a severe term of imprisonment, as the social costs of imprisonment (scaled with the number of offenders choosing imprisonment) will diverge. \\

In short, we see that if $r$ increases above some threshold (equation \ref{r_cond}) and $f$ decreases below some threshold (equation \ref{f_cond}) dependent on $r$, then theoretically speaking, then it is favorable to implement a severe term of imprisonment (see figure \ref{plotphase}), as the social costs of imprisonment will be contained. Note that in the case where $k_0$ is infinite, we essentially have a penal strategy where a fine is the effectively the only punishment option, as no rational agent would choose to be imprisoned infinitely long\footnote{There is the caveat here that in reality, the effective term of punishment can never be infinite as it is bounded above by the expected lifespan of the offender. Therefore, it may still be a possibility that the individual would choose imprisonment over fine if the fine is sufficiently large. However, given the constraint in \ref{f_cond}, we see that this is not a possibility in most practical cases, so the assumption of an infinite term of punishment is justified.}, so the optimization problem reduces to finding the optimal amount of fine and probability of apprehension, as the punishment of imprisonment is now inconsequential. 

\subsection{The Optimal Amount of Fine}
\label{final_opt}

The condition \ref{pq1} at equality can be interpreted as a curve in the $v-p$ plane, and the pair $\{v,p\}$ must be related as follow on the curve
\begin{equation}
\label{vp_curve}
v_c(p) = \beta\frac{b-\pi(p)s}{2\pi(p)r} 
\quad \text{or} \quad
p_c(v) = \pi^{-1}\big(\frac{b\beta}{2vr + s\beta}\big).
\end{equation}
If inequality \ref{r_cond} holds, then the region defined by condition \ref{pq1} corresponds to a region at the upper-right corner of the square $[0,1]\times[0,1]$ in the $v-p$ coordinate system. The region is bounded by the right and top edges of the square and the curve defined by equation \ref{vp_curve}. We then see that the smallest $v$ value in this region is at the intersection of the curve and the top edge, or $p=1$, which gives us
\begin{equation*}
v_{\min} = \beta \frac{b-s}{2r}.
\end{equation*}
Similarly, the smallest $p$ value is at the intersection of the curve and the right edge, or $v=1$, which gives us
\begin{equation*}
\pi(p_{\min}) = \frac{b\beta}{2r + s\beta}
\end{equation*}

Recall from section \ref{repar} that the optimization problem is to find the pair $\{v,p\}$ such that the following function is maximized
\begin{equation*}
\label{j}
J(v,p) =
l(1-v^{\alpha}) - \frac{\alpha}{\alpha-1}bw_m (1-v^{\alpha-1}) - pc_f (v^{\alpha} + \epsilon) - \frac{\alpha}{\alpha-1}pgsw_m( v^{\alpha-1} + \epsilon ) - c_p p
\end{equation*}
To have a better understanding of the structure of the problem, we write $J$ only with terms that depend only on $v$ and $p$ respectively and ignore any prefactors
\begin{equation}
\label{jvp}
\begin{split}
J(v) &\sim v^{\alpha-1}\Big\{ \frac{\alpha}{\alpha-1}(b-pgs)w_m - (l+pc_f) v \Big\} \\
J(p) &\sim -p\Big\{ c_p + c_f(v^{\alpha}+\epsilon) + \frac{\alpha}{\alpha-1}gsw_m(v^{\alpha-1} + \epsilon) \Big\}.
\end{split}
\end{equation}
We note that the term in the large bracket for $J(p)$ is always positive, so we see that for a fixed $v>v_{\min}$, the maximum of $J(p)$ is attained for the smallest possible value of $p$, or $p_c(v)$ (see equation \ref{vp_curve}). \\

The case for $J(v)$ is less trivial. Note that given some $p>p_{\min}$, we find that the local maximum of $J(v)$ is attained at
\begin{equation}
\label{vopo}
v_o(p) = \frac{(b - pgs)w_m}{l+p c_f}
\quad \text{or} \quad
p_o(v) = \frac{bw_m - lv}{gsw_m + c_fv},
\end{equation}
where $p_o$ is simply the inverse function of $v_o$. Note that $v_m(p) < \frac{bw_m}{l} < 1$\footnote{Note that It can be reasonably assumed that this value is smaller than $1$. Otherwise, we have $bw_m>l$, meaning that we have the trivial case where even for the poorest members, the utility gain returned from the offense is greater than the cost incurred on the victim. This means that the social cost of the crime would be far too low to justify any form of penal strategy.}. However, note that $v$ is bounded below by $v_c(p)$ (see \ref{vp_curve}), so we see that the global maximum is attained at (see appendix \ref{poly})
\begin{equation}
\label{v_opt}
\begin{split}
v^*(p) &= \max\big[ v_c(p), v_o(p) \big] \\
&= \max\big[ \frac{(b-pgs)w_m}{l+c_f p} \,,\, \beta\frac{b-\pi(p)s}{2\pi(p)r} \big] 
\end{split}
\end{equation}
or equivalently
\begin{equation*}
\boxed{
f^*(p) = \min\big[ \frac{2w_mr}{\beta} \,,\, (\frac{b-\pi(p)s}{b-pgs})(\frac{l+c_fp}{\pi(p)}) \big]
},
\end{equation*}
which gives us the optimal fine amount for some given probability of apprehension. Note that if we let $p=1$ for the sake of simplicity, we see that that the second argument of the $\min$ function is reduced to
\begin{equation*}
(l+c_f)(\frac{b-s}{b-gs})
\end{equation*}
meaning that the fine amount should scale positively with the loss incurred on the victim from the crime and the collection cost of the fine. This makes sense intuitively, if the victim's loss and collection cost is high, it is necessary to set the fine amount higher as well to reduce the number of offense thus the collection cost. \\

Perhaps it is less obvious why the fine amount should scale negatively with $g$, or the ratio between the total value of a member's output over his/her compensation (see section \ref{cost_pun}). To interpret this, note that for a large $g$ value, the offender is only evaluating a small fraction of the opportunity cost of imprisonment\footnote{This is because the offender is only seeing a small fraction of his productivity in his/her income, and fails to consider the other fraction which is converted to social utility gain.}, so in some sense, the deterrent effect of imprisonment is small relative to the social costs that it incurs. Therefore, imprisonment becomes a less cost-effective strategy in the sense that the ratio between the deterrent effect and its social cost is low, so it is necessary to set the amount of fine to be sufficiently small such that it becomes the effective strategy instead (as the majority of the population will choose a fine over imprisonment). 

\subsection{Optimal Probability of Apprehension}

Now that we have found the maximum over $v$ for every $p$, the optimization problem essentially reduces to an one variable optimization problem over only $p\in (p_{\min},1]$
\begin{equation*}
\boxed{
p^* = \text{argmax}_{p\in (p_{\min},1]} \big[ J(v^*(p),p) \big]
},
\end{equation*}
which can be done fairly quickly with any numerical solver. As expected, the optimization of the probability of apprehension presents the only non-trivial part of the optimization problem from a computational standpoint. Furthermore, we see that in order to solve this problem, it is necessary to specify the probability weighing factor $\gamma$ in the function $\pi(p)$, as the argument $v^*(p)$ (see equation \ref{v_opt}) depends explicitly on $\pi(p)$. 

\subsubsection{Special Case}

Note that under a certain condition, an analytic expression for the optimal $\{v,p\}$ pair is possible. In appendix \ref{two}, we show that if the following condition is satisfied
\begin{equation*}
p_o(1) \geq p_{\min}
\quad \text{or} \quad 
\frac{bw_m-l}{gsw_m+c_f} \geq \pi^{-1}\big( \frac{b\beta}{2r+s\beta} \big),
\end{equation*}
then the optimum is attained at
\begin{equation*}
v = 1 
\qquad 
\boxed{p = p_{\min} = \pi^{-1}( \frac{b\beta}{2r+s\beta} )}.
\end{equation*}
Note that $v=1$ implies that $w_0=w_m$, or
\begin{equation*}
\boxed{f = \frac{2rw_m}{\beta}},
\end{equation*}
which is exactly at the phase transition threshold (see equation \ref{f_cond}).

\section{Empirical Measurements}
\label{emp}

Note that the social welfare function is only defined if the distribution of the utility function parameters $\bsta$ is given for the population, otherwise there would be no knowledge of the expected numbers of offenders and non-offenders under any penal strategy (see section \ref{pop_part}), making the optimization impossible. In other words, for our model to have any practical application, it is necessary to device a method where the utility function parameters $\bsta$ can be accurately measured for any member, and the distribution parameters (see section \ref{dist_tr}) can be accurately estimated. \\

In this section, we discuss how the method of survey can be used as an effective tool to study empirically the distribution of $\bsta$ among the population. As a proof of concept, we created a short survey and distributed it among $207$ participants in mainland China through an online platform, {\it Wenjuanxing}, to ensure that we are uniformly sampling across all income brackets, age groups, and regions\footnote{It is of course very difficult to ensure that our samples are unbiased, as the distribution of population conditioned on having access to internet is by no means an accurate representation of the entire distribution. However, this is not a major problem if the parameters of interest is independent of the parameters biased by our sample selection. For example, we show in section \ref{inde} that the discount rate $k$ and level of wealth $w$ are distributed independently. This means that even if our sample does not faithfully capture the number of members in each income bracket, it does not necessarily mean that our sample of discount rates will be biased as well.}. Even though the sample size used in the study is relatively small, we were still able to obtain an accurate measurement of the distribution of the parameters $\bsta$ that our work requires. And in the future, we will extend this empirical study to capture a larger sample size, so that the distribution of $\bsta$ can be measured more accurately. \\

Note that the distribution of wealth $w$ should be well-determined (as the government is assumed to have a database containing the income of every member of the population), so the goal is to set up the survey questions to specifically measure the distribution of $\{k,\gamma\}$ (the discount rate and probability weighing factor), which are parameters that cannot be inferred from any database of information on the population. In section \ref{dist_dis}, we show that the distribution of $k$ in our sample in fact resembles the zero-inflated exponential distribution as assumed in section \ref{dist_tr}, and we report the confidence intervals of the two parameters $\{\beta,\rho\}$. Similarly, in section \ref{dist_prob}, we show that the distribution of $\gamma$ in our sample resembles a normal distribution, and we report the estimators of the mean and variance $\{ \mu_{\gamma}, \sigma_{\gamma} \}$ from the sample. Note that in addition to the error that enters when we estimate the distribution parameters from the samples, we also have to consider the error associated with the inaccurate responses of the participants. To take into account the errors at both stages, we constructed novel regression models which will be discussed in extensive details in appendices \ref{app_dis} and \ref{app_prob}. \\

In addition to the measurements of the $\{k,\gamma\}$ distributions. In section \ref{inde}, we show that the triple $\{w,k,\gamma\}$ are pairwise independent. The independence of $\{w,k\}$ is of particular importance as it is one of the core assumptions in the construction of our model which allows for the factorization of the pdf $f_{\{W,K\}}(w,k) = f_W(w)f_K(k)$ (see section \ref{dist_tr}). And in section \ref{harsh}, we show how the quantity $r$ (the harshness of imprisonment condition) can be measured within a population. The measurement of $r$ is crucial as it is one of the two parameters $\{f,r\}$ that induces a phase transition for the optimal penal strategy (see section \ref{sev_phase}). 

\subsection{Distribution of Discount Rates}
\label{dist_dis}

Recall from equation \ref{pdf2} that the distribution of discount rates among a population is assumed to be
\begin{equation*}
f_K(k) = (1-\rho) \delta(k)+\rho\frac{1}{\beta}\exp(-\frac{k}{\beta}).
\end{equation*}
We attempt to estimate the parameters $\rho$ and $\beta$ from the sample population by measuring the discount rate of each member. The technique is to provide the participant with a hypothetical scenario where he/she has been apprehended due to DWI and are forced to make a choice between an immediate punishment of a certain length and a delayed punishment of a longer length. The participant is then asked at least how long the punishment has to be delayed in order for him/her to choose the delayed punishment over the certain punishment. \\

Note that unlike previous studies \citep{hyper2}, a measurement of the participant's propensity, e.g. using the Likert scale \citep{likert}, towards a delayed punishment over an immediate punishment (where the parameters of both punishment options are fixed) does not suffice for the purpose of estimating the discount rate. In order to estimate the discount rate accurately, the exact point of indifference has to be measured between an immediate punishment of time $\tau_0$ and a delayed punishment of time $\tau_1$ delayed by $t$, where the participant him/herself has to specify at least one of the three parameters. It is decided that $\tau_0$ and $\tau_1$ should be fixed parameters while $t$ should be measured as the response. This results in the following natural formulation of the survey questions
\begin{equation}
\label{qk}
\begin{split}
\text{\it Given the choice between an immediate punishment of 2 hours and a delayed punishment of $\tau$,} \\
\text{\it I would prefer a delayed punishment if the delay is at least $t$. }
\end{split}
\end{equation}
where $\bs{\tau} = \{2.5,4,10,20\}$ is a fixed vector for every participant, corresponding four alternative terms of imprisonment. For every $\tau$, we request the participant to input a corresponding value of $t$, which gives us four values of $t$\footnote{Technically, the number of $t$ measurements is at most four. It is possible that the participant refuses to accept an imprisonment of length $\tau$ regardless of the delay, so it would be meaningless to measure $t$ for any imprisonment length above $\tau$. This may result in the number of $t$ measurements being less than $4$. See appendix \ref{app_dis} for a detailed discussion of how this is handled.}. Note that the disutility incurred on the offender from imprisonment, or $I(w,k,t,\tau)$, is specified in equation \ref{imp_dis}, so we can determine $k$ by setting the disutility of the delayed punishment equal to that of the immediate punishment of 2 hours
\begin{equation*}
I(w,k,0,2) = I(w,k,\tau,t) \implies t = \frac{1}{k}(\frac{\tau}{2}-1).
\end{equation*}
This results in an equation where $k$ can be uniquely specified given some pair of $\{t,\tau \}$ where $\tau>2$. Note that both $t$ and $\tau$ are measured in hours, so the unit of $k$ is $\text{hr}^{-1}$. \\

Of course, to perform regression analysis, we have to assume some error term in the above equation, and require more than one data points for accurate estimates of $\beta$ and $\rho$. Since in the survey the participant is given the option to specify $t$ either in days, weeks, months, or years, it is natural to assume that the error of $t$ scales with $t$ itself. In other words, a large value of $t$ gives rise to a greater uncertainty. Therefore, we should introduce a multiplicative error \citep{lognormal} associated with $t$, or
\begin{equation}
\label{tautre}
t(1+\epsilon) = \frac{1}{k}(\frac{\tau}{2}-1),
\end{equation}
where $\epsilon$ is assumed to be a normal random variable (RV) of zero mean. This is a rather unusual model\footnote{The regression model that we use here is novel, as we assume some multiplicative error term that is {\it not} log-normal. This gives rise to an interesting relationship between the standard error of the estimator and the estimated error term variance. A complete treatment of this topic is given in appendix \ref{app_dis}.}, and it presents some difficulties in performing error analysis, which we discuss in appendix \ref{app_dis}.  \\

For now, we simply state the unbiased estimator \citep{regress} of $k$ for a participant (which we prove to be in fact unbiased in appendix \ref{app_dis}). If we assume that participant $i$ has a discount rate of $k_i$, and responded with a minimal delay of $t_{ij}$ to question corresponding to the imprisonment length $\tau_{j}$ (see question \ref{qk}), then the unbiased estimator of $k$ is given by the following average of ratios
\begin{equation*}
\hat{k}_i = \frac{1}{m}\sum_{j=1}^m \frac{\frac{\tau_j}{2}-1}{t_{ij}},
\end{equation*}
where $m$ is simply the number of imprisonment options. Note that in the case where the participant responded with $t_{ij}=\infty$ corresponding to some value of $\tau_j$ (meaning that he/she will not choose an imprisonment of length $\tau_j$ under any condition of delay), then we can naturally treat this ratio as being $0$. The standard error for the estimator of $\hat{k}_i$ is rather complicated, and we will discuss in greater detail in appendix \ref{app_dis}. \\

Assume now that the number of participants is $n$, then we have a sample of $\mbf{k}=\{k_1,k_2,...,k_n\}$ measured from the participants from which we can use to estimate the $\rho$ and $\beta$ parameters of the zero-inflated exponential distribution (see equation \ref{pdf2}). The natural thought would be to use the fraction of non-zero $k$ values as an estimator of $\rho$, and the mean of these values as an estimator for $\beta$, or
\begin{equation*}
\hat{\rho} = \frac{1}{n} \sum_{i=1}^n \mbf{1}(k_i \neq 0)
\qquad
\hat{\beta} = \big[ \sum_{i=1}^n \mbf{1}(k_i \neq 0) \big]^{-1} \sum_{i=1}^n \hat{k}_i,
\end{equation*}
where $\mbf{1}$ denotes the indicator function \citep{real} which evaluates to $1$ if its argument is true and $0$ otherwise. We show in appendix \ref{app_dis} that these estimators are in fact unbiased, and provide a discussion on constructing the confidence interval for the two estimators. \\

We provide here an intuitive discussion of the origin of the difficulty of the regression analysis. The main point to note here is that the error in the constructed distributed is propagated in two stages. At the first stage, the error comes from the response of the participant (see equation \ref{tautre}), resulting in an error attached to the estimator $\hat{k}_i$ for each participant $i$. At the second stage, when we use all the values of $\hat{k}$ measured from the entire sample to construct the probability distribtion $f_K(k)$, there must also be errors associated with the distribution parameter estimators $\{\hat{\beta},\hat{\rho}\}$, just simply as a result of sampling statistics \citep{probability}. In the regression analysis, the errors in two stages are convoluted, which is the main source of complexity for performing error analysis. \\

\begin{figure}
\centering
\includegraphics[scale=0.6]{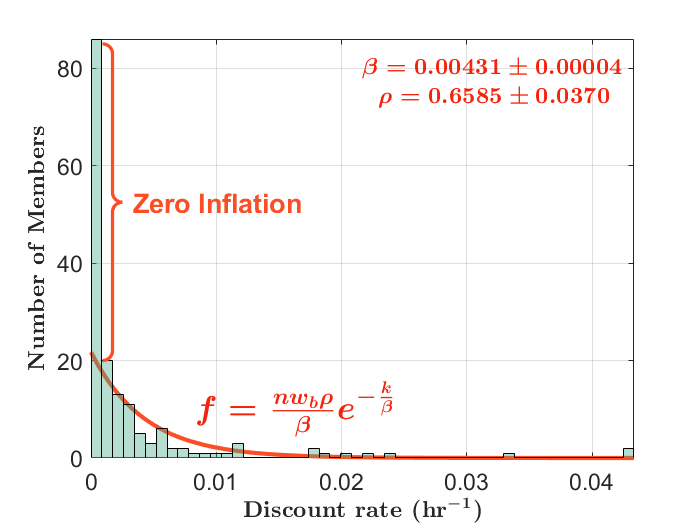}
\caption{\label{kplot} This histogram shows the distribution of probability weighing factors among a population sample of size $n=164$. The bin width of the histogram is $w_b = 0.02$. Note that there is a large concentration of zero discount rates, and the distribution of non-zero discount rates is well fitted with an exponential function (scaled appropriated with the bin width and number of samples). The fraction of non-zero discount rates is estimated to be $\rho = 0.6585\pm 0.0370$, and the mean discount rate estimated from the non-zero samples is $\beta = 0.00431\pm 0.00004\,(\text{hr}^{-1})$. The uncertainty corresponds to an one-sigma interval, the derivation of the standard error is given in appendix \ref{app_dis}. Color in print.}
\end{figure}

The distribution of the discount rates is presented as a histogram in figure \ref{kplot}, with the $\beta$ and $\rho$ estimators reported as
\begin{equation*}
\hat{\beta} = 0.00431\pm 0.00004\,\text{hr}^{-1}
\qquad
\hat{\rho} = 0.6587\pm 0.0370
\end{equation*}
The distribution is sampled from 164 out of the 207 participants, where the participants who chose not to answer this section of questions are excluded. Note that the distribution is dominant near zero, and the tail of the distribution is well fitted by an exponential distribution, which justifies the assumption that $f_K(k)$ follows a zero-inflated exponential distribution (see section \ref{dist_tr}).

\subsection{Probability Weighing Factor}
\label{dist_prob}

Recall in equation \ref{pi} that the probability weighing function can be expressed as follow
\begin{equation*}
\pi(p) = \frac{p^{\gamma}}{(p^{\gamma} + (1-p)^{\gamma})^{1/\gamma}},
\end{equation*}
where the probability weighing factor $\gamma$ is assumed constant for all members of the population. In reality, this assumption is not accurate as the probability weighing factor $\gamma$ actually varies among the members of the population\footnote{Note that this fact does not affect the previous analytic results of the optimal penal strategy, under the condition that the distribution of $\gamma$ and $\{w,k\}$ are roughly independent.}. In fact, we show that the distribution of $\gamma$ among a population follows closely a normal distribution. \\

The probability weighing factor $\gamma$ is measured using the concept of {\it certainty equivalent} \citep{risk}. Imagine the scenario where an individual is given a choice between an uncertain reward with some probability of being realized and a guaranteed reward. If the individual is indifferent between the two rewards, then the guaranteed reward is denoted as the certainty equivalent of the uncertain reward. In the context of deterrence, an individual would be indifferent towards committing the crime and not committing the crime if the net utility gain from the crime is exactly zero, or
\begin{equation}
\label{cer_equi}
B - \pi(p)\Big[ S + D \Big] = 0
\end{equation}
where $S$ denotes the stigma associated with apprehension and $D$ is the disutility of the punishment (see section \ref{gen}). Therefore, given some probability of apprehension, we say that the certainty equivalent of committing the offense (and facing the risk of being caught) is to give up on the offense and forfeit the potential reward of amount $B$. To see how the probability weighing factor can be measured with the concept of certainty equivalent, we first rewrite equation \ref{cer_equi} as follow
\begin{equation}
\label{pratio}
\pi(p) = \frac{p^{\gamma}}{(p^{\gamma} + (1-p)^{\gamma})^{1/\gamma}} = \frac{B}{S+D}.
\end{equation}

To determine the value of $\gamma$ for a participant, we have to measure one of the three parameters, $\{B,S,D\}$, corresponding to different values of $p$. For example, to determine $\gamma$ uniquely, one possible option is to make at least three measurements of $D$ corresponding to three different values of $p$\footnote{The reason three data points is required is because there are three unknowns in the system of equations $\{\gamma,B,S\}$. Of course, this is only a necessarily condition to solve for $\gamma$, as there are cases where the system of equations gives no solutions (such as when the measured values of $D$ somehow increase with increasing $p$, which violates the property of $\pi(p)$ being a monotonously increasing function.)} \citep{nonlinear}. To perform the measurements, we can place the participant in a scenario where he/she may be tempted to commit the offense under some level of apprehension probability $p$. Then the participant is asked to respond with the amount of fine that is sufficient to deter him/her from committing the offense\footnote{Note that theoretically speaking, this can also be done the other way. In other words, we can ask the participant given some fine amount, what level of apprehension probability is sufficient to deter him/her from committing the crime. However, note that giving a response in terms of probability is usually much harder than giving a response in a dollar amount, with the latter being a much more ubiquitous object in daily life.} (thus giving us a measure of $D$). Note that if we choose to make the measurements on $D$, then we have three unknowns to determine, $\{\gamma,B,S\}$, which is slightly problematic for non-linear regression analysis (see appendix \ref{app_prob}) as there are three regression coefficients to estimate, and this requires many data points to achieve sufficient accuracy \citep{nonlin_re}. \\

To reduce the degree of freedom in the problem, we note that the number of unknown coefficients can be reduced to two if we realize that $S$ and $D$ appear in additive form on the right hand side in equation \ref{pratio} (and appear no where else). This means that the two coefficients can be effectively combined into one, or simply $S+D$, so an alternative option is to not measure $S$ or $D$ at all and perform measurements on $B$ instead. Of course, it is very difficult to measure $B$ under regular condition as there seems to be no way to have the participant quantify the utility gain from the DWI offense. As an example, consider the case where an offender commits DWI by drunk driving back home late night for the purpose of not wanting to miss work tomorrow, the utility gain of ``not missing work" (or the opportuntiy cost of missing work) is obviously difficult to quantify, as it depends on many factors that cannot be controlled by the survey (such as how lenient his/her boss is). \\

Fortunately, under the unique technological context of mainland China, there is a clever way which we can convert this utility gain into an exact dollar amount. To be more specific, in mainland China, there is a prevalent mobile application, {\it Didi}, which allows the user to request a designated driver with a very short wait time. This makes the measurement of $B$ much easier for us as it allows us to set the utility gain from the crime equal to the amount of money that would be saved from not requesting the Didi service\footnote{This is of course under the assumption that for the participant, there is no intrinsic value to the activity of performing DWI itself. In other words, we are assuming that the participant does not receive any form of psychological stimulation from performing DWI, meaning that he/she would not choose to commit DWI ``for no reason". Of course, this assumption breaks down in reality, as there exists offenders who derive pleasure from performing DWI, and this psychology constitutes a major factor in their evaluation of utility function. However, in this study, we find that only 2 out of the 207 participants responded that they derive pleasure from DWI, so for the purpose of our empirical analysis, we choose to ignore this fraction of population.} (or the opportunity cost of not committing DWI). In other words, we say that the certainty equivalent of facing the risk of apprehension by committing DWI is to instead spend an amount of $B$ on requesting a designated driver. To measure this certainty equivalent, we formulate the survey question as follow
\begin{equation}
\begin{split}
\text{\it Given that the probability of apprehension is p and the fine is 500 RMB\footnote{RMB is short for Renminbi, or Chinese Yuan, which is the official currency of mainland China.},}  \\
\text{\it I would choose to request a designated driver if the cost does not exceed \_\_\_\_ RMB.}
\end{split}
\end{equation}
where the participant is asked to fill in the blank with a dollar amount based on his/her preference. Note that in this hypothetical scenario, we also specified that the driver is guaranteed to arrive under 2 minutes, as to minimize the opportunity cost associated with the wait time, so that the uncertainty in the measurement of $B$ is reduced. Recall that by directly measuring $B$, the number of regression coefficients decreases from $3$ to $2$, with the $2$ coefficients being $\{S+D,\gamma\}$ (see appendix \ref{app_prob}).\\

\begin{figure}
\centering
\includegraphics[scale=0.6]{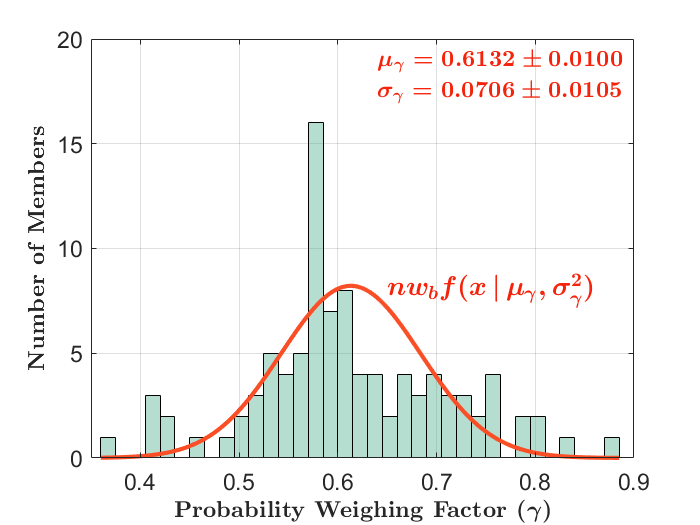}
\caption{\label{gammaplot} This histogram shows the distribution of discount rates among a population sample of size $n=97$. The bin width of the histogram is $w_b = 0.02$. The distribution of the probability weighing factor is fitted under a normal distribution (appropriately scaled with the number of samples and bin width). The mean of the distribution is estimated to be $\mu_{\gamma} = 0.6132\pm 0.0100$, and the standard deviation of the distribution is estimated to be $\sigma_{\gamma} = 0.0706\pm 0.0105$. The uncertainties of the estimators correspond to their one-sigma intervals. See appendix \ref{app_dis} for a discussion of how the confidence intervals of the estimators are constructed. Color in print.}
\end{figure}

The values of $p$ used in constructing the hypothetical scenarios are fixed by the vector,
\begin{equation}
\mbf{p} = \{0.05,0.1,0.25,0.5,0.75,0.9,0.95,0.98,1\},
\end{equation}
for every participant. We see that the probability of apprehension ranges from unlikely (or $5\%$) to certain (or $100\%$). For participant $i$, if he/she responds with $B_{ij}$ for a given probability of apprehension $p_j$, we then construct the following model relating $B_{ij}$ to $p_j$
\begin{equation}
\label{gre}
B_{ij} - \pi(p_j,\gamma_i) (S_i + D_i) = \epsilon_{ij},
\end{equation}
where $\epsilon_{ij}$ is an independent normal error term with mean $0$ and variance $\sigma_i^2$, with the variance being a variable that depends on the participant. This model presents a non-linear regression problem where the goal is to find an unbiased estimator $\hat{\gamma}_i$ for the probability weighing factor of the individual, and construct its corresponding confidence interval (see appendix \ref{app_dis} for a detailed discussion of the regression model). Using the estimators of $\gamma$ evaluated for all participants, we then construct the distribution of $\gamma$ values among the population.  We assume that the distribution is normal with mean $\mu_{\gamma}$ and variance $\sigma_{\gamma}^2$, with the explicit expressions for their estimators and standard errors given in appendix \ref{app_prob}. The distribution of $\gamma$ is presented as a histogram in figure \ref{gammaplot}, with the $\mu_{\gamma}$ and $\sigma_{\gamma}$ estimators reported as
\begin{equation*}
\mu_{\gamma} = 0.6132\pm 0.0100
\qquad
\sigma_{\gamma} = 0.0706\pm 0.0105.
\end{equation*}
Note that similarly to the discussion in section \ref{dist_dis}, this model presents much difficulty in regression analysis as the error is propagated in two stages. \\

It is interesting to note that since $S_i+D_i$ enters as a regression coefficient in equation \ref{gre}, the value of $S_i+D_i$ can be estimated for each member, and we can compute the value of $S_i$ as a ``by-product". To find the value of $S$, we assume $D = 500\,\text{RMB}$ to be the utility loss of each member from the fine, and simply subtract it off the estimator of $S_i+D_i$. Since the value of $S_i$ is not of major interest to this study, we simply report the median value of the estimators of $S$ from the samples, which is $\tilde{S} = 68.2\,\text{RMB}$. 

\subsection{Pairwise Independence of $\bs{\{w,k,\gamma\}}$}
\label{inde}

In section \ref{dist_tr}, a major assumption that we made is that the wealth level and discount rate, $\{w,k\}$, are distributed independently among a population. In this section, we extend this assumption further and show empirically that the triplet $\{w,k,\gamma\}$ are in fact pairwise independent\footnote{Theoretically speaking, pairwise independence does not imply mutual independence \citep{probability}, but this distinction is usually very subtle in most practical scenarios.}. To show the independence of two variables, we split the samples at the median of one of the variables and show that the distributions of the other variable in the two groups agree within uncertainty. Note that in this analysis, $w$ is quantified by the reported income of the participant\footnote{Assuming that the actual level of wealth is some function of an individual's salary. Then for a population, showing the independence of the distribution of some behavioral traits with respect to salary also implies independence with respect to the level of wealth. This is because functions of two independent RVs are also independent RVs themselves \citep{probability}.}.\\

To be more specific, consider that we have $n$ samples from RVs $\mbf{A}=\{A_1,...,A_n\}$ and $\mbf{B}=\{B_1,...,B_n\}$ (where $A_i$ and $B_i$ correspond to the same sample). We can reorder the samples by performing some permutation on the indices \citep{group} such that the samples are split at the median of $A$\footnote{The reason why we choose the median is because we wish to split the data points into two groups, such that the errors of the estimators (which we assume to scale inversely with respect to the square root of the sample size, or $\sim \frac{1}{\sqrt{n}}$) are in some sense ``distributed equally" between the two groups, so that there are no excessive errors in any of the groups. This is for a more accurate analysis and is not strictly necessary. Note that however, this is {\it not} a form of median test \citep{median}, as the median is {\it not} used to characterize the distributions of the two groups but serves rather as a grouping condition of the samples.}. To be more specific, if we denote the permutation as $\sigma$, then the following is require
\begin{equation*}
\forall i \leq \lfloor \frac{n}{2} \rfloor\,\big[ A_{\sigma(i)} < \tilde{A} \big],
\end{equation*}
where $\tilde{A}$ denotes the median of the samples $\mbf{A}$. We can then split the samples $\mbf{B}$ into two groups
\begin{equation*}
\mbf{B_{\alpha}} = \{ B_{\sigma(1)},...,B_{\sigma(\lfloor \frac{n}{2} \rfloor)} \}
\qquad
\mbf{B_{\beta}} = \{B_{\sigma(\lfloor \frac{n}{2} \rfloor)},...,B_{\sigma(n)}\},
\end{equation*}
where the elements in the two samples may not be exactly equal. In particular, it is possible for $|\mbf{B_{\beta}}| > |\mbf{B_{\alpha}}|$ if $n$ is odd. For the sake of the example, let's say we know that $B$ is normally distributed (with mean $\mu_B$ and $\sigma_B$) and the goal is to show that it is independent of $A$. We then find the sample means and variances of the two groups $\mbf{B_{\alpha}}$ and $\mbf{B_{\beta}}$, $\{\hat{\mu}_{\alpha},\hat{\sigma}_{\alpha}\}$ and $\{\hat{\mu}_{\beta},\hat{\sigma}_{\beta}\}$, and show that they agree respectively between the two groups within their confidence intervals \citep{moment}. The analysis can also be done in the other direction (splitting data points at the median of $B$ and compare of the sample means and variances of $A$ for the two groups). Note that this procedure is rather general, and should apply not only to normally distributed RVs but also other distributions as well\footnote{In some sense, this is similar to an ANOVA test \citep{anova} between two variables (where normality is not assumed) where we essentially ``transform" one of the variables into a categorical variable where the categorization condition for the sample is based on its location relative to the sample median. Unlike ANOVA, however, the goal is not only reject the hypothesis that the means of the two groups are different, but the hypothesis that {\it any} of the parameters characterizing the underlying distribution is different, which is usually a much stronger condition. }, assuming that the distribution is well-behaved. In some sense, the goal is to retain the null hypothesis that the distribution parameters of the two groups are the same, hence implying that $A$ and $B$ are independent RVs. \\

\begin{figure}
\centering
\includegraphics[scale=0.4]{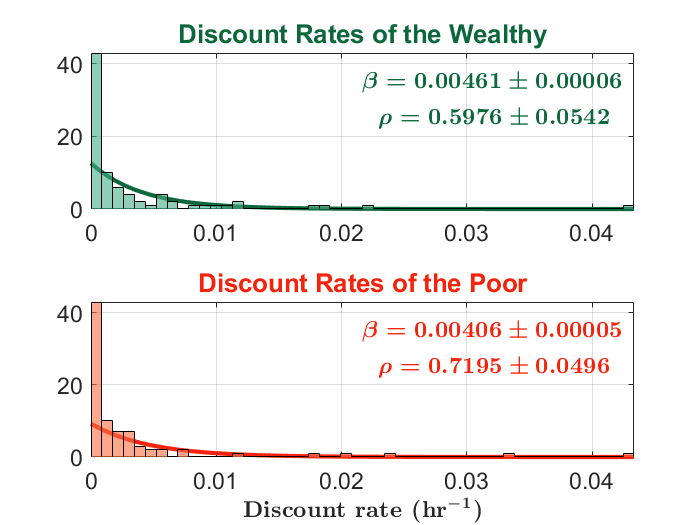}
\includegraphics[scale=0.4]{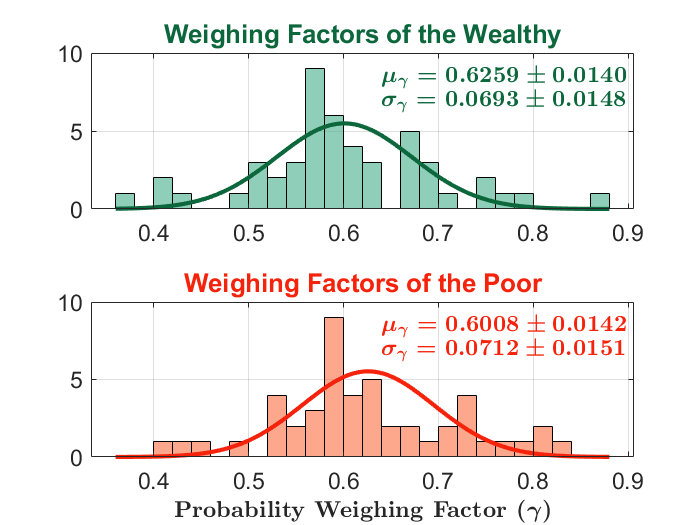}
\includegraphics[scale=0.4]{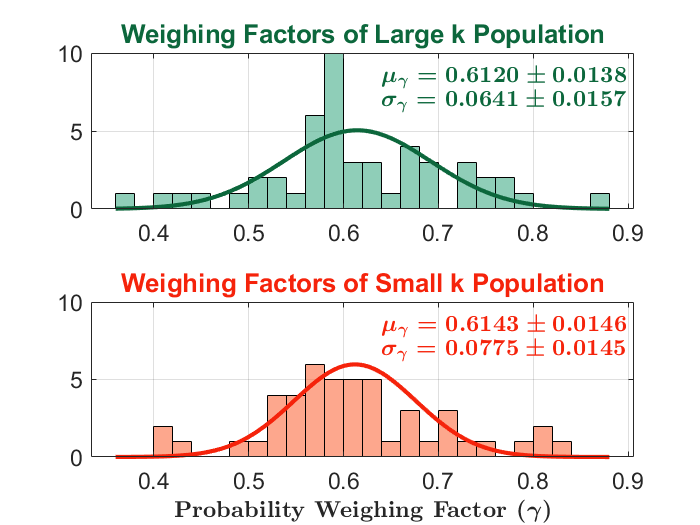}
\caption{\label{sal} Top left figure: the distributions of the discount rates among two groups split at the median salary of the sample. Top right figure: the distribution of the probability weighing factors among two groups split at the median salary of the sample. Bottom figure: the distribution of the probability weighing factors among two groups split at the median discount rate of the sample. Note that in all three cases, the distribution parameters of the two groups (reported on the corresponding histograms) agree well with each other. The largest difference is the $\hat{\rho}$ values (reported in the top left figure) of the $k$ distributions between the wealthy and poor, corresponding to a $z$ value of $1.66$. Color in print.}
\end{figure}

To show the pairwise independence of $\{w,k,\gamma\}$, we have to check the three possible pairings of the parameters separately. To check the independence of $\{w,k\}$, we split the samples at the median salary of the sample, and show the distributions of $k$ among the ``wealthy" and the ``poor" are the same (see the top left histogram in figure \ref{sal}). Note that the difference between the two $\rho$ values corresponds to a $z$ value of $1.66$, which is still small enough for the null hypothesis to be retained\footnote{By some standards, this $z$ value is actually large enough for one to reject the null hypothesis that the distributions of $k$ in the two groups are equal. However, even if we assume that $\rho$ is dependent on the level of wealth, this does not matter too much for the analysis carried out in section \ref{sev_phase}, as the asymptotic behavior of the social welfare function for large $k$ is governed by the mean of the exponential distribution $f_K(k)$, or $\frac{1}{\beta}$, and we see that the values of $\beta$ agree very well between the two groups. } \citep{hypo}. Similarly, to check the independence of $\{w,\gamma\}$, we again split the samples at the median salary, and show the distributions of $\gamma$ among the two groups are the same (see the top right histogram in figure \ref{sal}). Finally, to check the independence between $\{k,\gamma\}$, we split the samples at the median $k$ value, and show the distributions of $\gamma$ among the two groups are the same (see bottom histogram of figure \ref{sal}).\\

It should be noted that we did not opt to perform standard hypothesis rejection techniques \citep{hypo} which is rather difficult to be extended to account for two-stage errors (see the discussion in section \ref{dist_dis}). Instead, it makes much more sense to just compare the distribution parameters directly between the two groups, as the method for constructing the confidence intervals of these distribution parameters has already been developed in the previous two sections (section \ref{dist_dis} and \ref{dist_prob}). 

\subsection{Measuring Harshness of Imprisonment}
\label{harsh}

In equation \ref{f_cond}, we express the phase transition threshold as
\begin{equation*}
f < \frac{2rw_m}{\beta}.
\end{equation*}
Therefore if we can somehow measure the value of $r$ for a population, then we have effectively determined the upper bound of $f$ for the optimal penal strategy. To measure $r$ implies that we need to quantity how the population perceives the imprisonment condition. At first, this seems to be a rather intractable task as there is no obvious way for the participant to express the harshness of the imprisonment condition in terms of some well defined utility; however, we see that it is not necessary for the participant to quantify $r$ directly if we can map the imprisonment punishment to some other punishment option which we can attach an exact dollar figure to. \\

Therefore, the natural thought would be to locate the point of indifference between a fine and imprisonment for each member. Recall from equation \ref{curve} that the point of indifference can be expressed as
\begin{equation}
\label{r}
r = \frac{kf}{w\log[ 1+\frac{k\tau}{1+kt} ]}.
\end{equation}
This equation means that when a member is given the choice between a fine of $f$ and an imprisonment of length $\tau$ delayed by $t$, the member would be indifferent between the two punishments. Note that for every member of the population, the discount rate $k$ can be measured using the method described in section \ref{dist_dis}, and the level of wealth can be inferred from the reported salary figure. In the case where $k=0$ for a member, we can take the limit of equation \ref{r} as $k\to 0$
\begin{equation*}
\lim_{k\to 0} \frac{kf}{\log\big[ 1+\frac{k\tau}{1+kt} \big]} = \frac{f}{\tau}.
\end{equation*}
Now, the task that still remains is to determine the triple $\{f,t,\tau\}$. \\

This is done indirectly through the following pair of survey questions:
\begin{equation}
\begin{split}
&\text{\it {\bf 1}. Given that the probability of apprehension is $0.02\%$}\\
&\text{\it and the punishment is a fine of 500 RMB,} \\
&\text{\it I would request a designated driver if the cost does not exceed \_\_\_\_ RMB.} \\
\\
&\text{\it {\bf 2}. Given that the probability of apprehension is $0.02\%$} \\
&\text{\it and the punishment is an immediate detention.} \\
&\text{\it If the cost for requesting a designated driver is (the response to Q1),} \\
&\text{\it I would request a designated driver if the length of detention is greater than \_\_\_\_ hours. }
\end{split}
\end{equation}
Note that the first question is given as one of the questions used to measure the participant's probability weighing factor (see section \ref{dist_prob}). To see how this pair of questions locates the point of indifference for the member, consider the following equation system
\begin{equation}
\label{trip}
\begin{cases}
B - \pi(p)\big[ f+S \big] = 0 \\
B - \pi(p)\big[ I(w,k,t,\tau)+S \big] = 0,
\end{cases}
\end{equation}
where $S$ is the stigma associated with apprehension (which we can assume to be unknown even though it can be easily measured as shown in section \ref{dist_prob}). The first equation corresponds to the first question, where $f$ is given to be 500 RMB, and $B$ is the utility gain from the crime, which as discussed in section \ref{dist_prob}, is simply the price of requesting a designated driver. And this dollar amount is given as the participant's response in the first question. The value of $B$ is then carried over to the second question, as the response from the first question is used as part of the question statement for the second question. From the second question, we can obtain the value of $\tau$ as the maximum length of imprisonment that the member can tolerate (while setting $t=0$ as the punishment is immediate). Note that the two equations implies
\begin{equation}
\label{doub}
f = I(w,k,0,\tau),
\end{equation}
which is simply the equation for the point of indifference as appeared in equation \ref{r}\footnote{Note that equation \ref{trip} implies equation \ref{doub}, but the reverse is not true. A clear way to see this is to note that the solution to equation \ref{trip} is the point of indifference for {\it three} options: committing the crime, accepting a fine, and accepting. It corresponds to the point where all three subsets $\{\Omega_0,\Omega_1,\Omega_2\}$ meet (see figure \ref{plot_part}). On the other hand, the solution to eqution \ref{doub} is simply the point of indifference for {\it two} options: accepting a fine and accepting being imprisoned, but it is {\it not} the point of indifference between committing the crime or not. In other words, this means that the RHSs of the two equations in \ref{trip} do not have to equate to zero, meaning that $B$ doesn't have to be set as the certainty equivalent of committing the crime.}. In some sense, we are locating the point of indifference indirectly through two separate scenarios where the two punishment options are equated to a third quantity $B$, or the utility gain from the crime. And equating the two punishment options to a third quantity guarantees the equality of the two options themselves\footnote{Note there are two reasons why we do not equate the two punishments directly by asking a question such as: {\it Given the punishment options of a fine of amount $f$, I would prefer to be imprisoned instead if its length does not exceed $\_\_\_\_$ hours}. The first reason is that we need to have some structural consistency in the survey questions so that the participant does not become easily confused by any structural disorder, and asking a question where two punishment options are directly given (instead of just one) necessarily breaks this structural flow. The second reason is that by assuming that the offender is already apprehended and forced to choose one of the two punishment options, we are constructing a scenario that may be {\it unrealistic} to the participant, so the response given may not be a faithful representation of his/her behavioral traits. }. \\

Going back to equation \ref{r}, we see that $f=500\,\text{RMB}$ is already given as a fixed amount in the questions, $t=0$ is set zero as the punishment is immediate, and $\tau$ is given as the participant's response to the second question. This allows us to determine $r$ uniquely if we assume $w$ to be the reported salary of the participant. Then, we can solve for $r$
\begin{equation*}
r = 
\begin{cases}
kf/\big[ w\log ( 1+k\tau )\big]\quad &\text{if}\,k>0 \\
f/(w\tau) &\text{if}\,k=0,
\end{cases}
\end{equation*}
noting that the unit of $r$ is in $\text{hr}^{-1}$. We measure $r$ using this method for every member, and find the median value of $r$ of the sample to be
\begin{equation*}
\tilde{r} = 0.0505\,\text{hr}^{-1}.
\end{equation*}
This result, combined with the estimate of the mean discount rate $\hat{\beta}=0.00431$ in section \ref{dist_dis}, allows us to compute the upper bound of the fine amount to be
\begin{equation*}
f < \frac{2rw_m}{\beta} = w_m(2\times\frac{0.0505}{0.00431}) \approx 23w_m.
\end{equation*}
In other words, from this empirical analysis, we see that the amount of fine must be below 23 times the minimum income in the population in order for a severe penal strategy to be effective. One can see that this condition is usually satisfied in most of the penal strategies that are currently being implemented.

\section{Extensions}
\label{ext}

Up to this point, the main theoretical framework and empirical analysis have been fully discussed. We here provide a brief discussion of some further generalizations and possible extensions that can be made to our study, and give a few examples of how other classes of crimes (other than DWI) can be effectively studied under this model.

\subsection{Utility Function Factorization}
\label{fac}

Our model of social welfare maximization is dependent crucially on knowing how the parameters of the utility function are distributed among a population. It is well known that given a full utility function, there are multiple decompositions possible \citep{factor}. In particular, there should be a particular decomposition of the full utility function into multiple independent parameters which can be studied. In this work, we focused on the three parameters $\{w,k,\gamma\}$ (wealth, discount rate, and probability weighing factor) as we believe they are extremely crucial components of the utility function while being relatively easy to measure among a population. Even though we have shown that the three parameters are distributed independently, there is no reason to assume that these parameters form a complete basis \citep{linalg} for describing the full utility function. In other words, for a finer description of the utility function, we have to explore the contributions of other factors such as genetics, society, and religion. Not only that, but an investigation on {\it how} the distribution these parameters among a population can be measured is crucial. 

\subsection{Dynamic Criminal Activities}
\label{dy}

Furthermore, in section \ref{gen}, our model assumes only two-body interactions of crimes, and that the appearance of opportunities on each edge is an independent Poisson process with constant rate. There are multiple generalizations that can be made for the model to capture more realistic and interesting scenarios. First, the most natural generalization is to have to model account for crimes that involve more than two parties (such as arson), which can be described formally as a hypergraph\footnote{A hypergraph is a graph whose edges can connect any number of vertices.} \citep{hype_graph}. \\

Second, to capture the effect of distance in criminal activities, we have to note that an opportunity is more likely going to arise in areas close to the offender than in areas far from the offender \citep{crime_dis}. We can then define naturally an Euclidean graph \citep{geo_graph} with the vertices corresponding to the rough locations of the members of the population, and the length of an edge corresponding to the distance between the two member. We can then set the rate at which criminal opportunities arise between two members to decrease with distance. For example, we can set the rate $\lambda$ to decay exponentially with distance, or 
\begin{equation*}
\lambda_{\zeta}(\omega_i,\omega_j) = e^{-ad(\omega_i,\omega_j)},
\end{equation*}
where $a$ is some decay factor, and $d(\omega_i,\omega_j)$ denotes the distance between member $\omega_i$ and $\omega_j$. \\

Lastly, it is also possible to model the rate $\lambda$ as a dynamic variable which varies in time. This can be used, for instance, to capture the phenomenon of revictimization, where a victim of some particular crime in the past is more likely to be victimized again in the future \citep{revic}. If we denote $N(\zeta,\omega_j)$ as the number of times that the member $\omega_j$ has been a victim of crime $\zeta$ in the past, then we can, for example, model the rate of opportunities connecting from any member $\omega_i$ to $\omega_j$ to scale logarithmically\footnote{Note that we don't assume a linear scaling as this would imply an exponential growth in criminal activities, which is clearly not realistic.} with $N(\zeta,\omega_j)$, or
\begin{equation*}
\lambda_{\zeta}(\omega_i,\omega_j) = a + b\log\big[ N(\zeta,\omega_j) + c \big],
\end{equation*}
Of course, it is possible to make the model as convoluted as we can, but this comes at the expense of analytic tractability. To achieve a balance between the two, the model should be complicated enough to realistically capture the crime of interest, but also simple enough such that non-trivial properties can be derived. 

\subsection{Burglary}
\label{bur}

In section \ref{ass}, we made the simplifying assumption that the utility gain from the crime for the offender scales proportionally with his/her wealth level, which results in a simple partition of the population into three groups (see figure \ref{plot_part}). This assumption is only particular to the crime of DWI, where it is assumed that the motivation behind committing DWI is the opportunity cost of otherwise not doing so, and the opportunity cost scales proportionally with the offender's level of wealth. This assumption is not necessarily justified for all classes of crimes. A simple example would be the crime of burglary, where the utility gain from the crime is independent of the level of wealth of the offender (as the dollar value of items that the offender obtains from the crime should not depend on how rich the offender is\footnote{However, if we assume that the utility function scales logarithmically with the amount of money that an individual has \citep{uti_w}, then the marginal utility gain of a dollar for an individual scales inversely with his/her current level of wealth, in contrast to the DWI crime where the utility gain from the crime scales proportionally with the offender's level of wealth.}). \\

For the sake of simplicity, let's assume that the utility gain is a constant $B$, and the probability of apprehension is 1, then the utility function of the individual can be written as (refer to equation \ref{det})
\begin{equation*}
B - \min\big[ f,I(w,k,t,\tau) \big].
\end{equation*} 
One can immediately note that when $f<B$ (or if the amount of fine does not exceed the value of items that the offender can potentially obtain from the crime), then the utility function will be positive, and an offense will always be realized whenever an opportunity is presented, so the partition $\Omega_0$ will always be empty. Similar to the case where $w_0<w_m$ for a penal strategy targeting DWI (see section \ref{pop_part}), this results in a strategy that only incurs social costs but does not result in any benefit in the form of criminal deterrence.\\

Let's then only consider the non-trivial case where $f>B$, then in order for a member to be deterred, his/her utility gain from the crime must be negative, or
\begin{equation*}
B < \min\big[ f,I(w,k,t,\tau) \big]
\quad \implies \quad
\big\{B < f\big\}\,\land\,\big\{ k < \frac{rw}{B}\log\big[ 1+\frac{k\tau}{1+kt} \big] \big\},
\end{equation*}
where we used the expression for $I(w,k,t,\tau)$ as appeared in equation \ref{imp_dis}. Note that the first condition is automatically satisfied by assumption, so we focus on the second condition, which can be approximated as (see appendix \ref{asy})
\begin{equation}
\label{kb}
k < \frac{rw}{B} \log\big[ 1+\frac{\tau}{t} \big],
\end{equation}
meaning that the inequality represents the region below a straight line with slope $\log\big[ 1+\frac{\tau}{t} \big]$ in the $w-k$ coordinate system, which corresponds to the subset $\Omega_0$. Similar to the discussion in section \ref{pop_part}, we see that the complement of this subset $\Omega/\Omega_0$ (corresponding to the subset of offenders) can be further partitioned into two subsets based on whether the offender chooses a fine or imprisonment. And this partition can be represented as the same {\it partition curve} as defined in equation \ref{curve}
\begin{equation}
\label{kf}
k = \frac{rw}{f}\log\big[ 1+\frac{\tau}{t} \big].
\end{equation}
Note that $f<B$, so the slope of this line is steeper than that of the line in equation \ref{kb}. In some sense, the line defined in equation \ref{kf} is always ``higher" than the line defined in equation \ref{kb}. We then see that the subset $\Omega_1$ corresponds to the region between the two lines, which can be described by the following inequality
\begin{equation}
\frac{rw}{B}\log\big[ 1+\frac{\tau}{t} \big] < k < \frac{rw}{f}\log\big[ 1+\frac{\tau}{t} \big].
\end{equation}
And the subset $\Omega_2$ corresponds to the region above the higher of the two lines, which can be described by the following inequality
\begin{equation}
k > \frac{rw}{f}\log\big[ 1+\frac{\tau}{t} \big].
\end{equation}
To summarize, the partitioning of the population into the subsets $\{\Omega_0,\Omega_1,\Omega_2\}$ is defined by two (non-parallel) lines in the $w-k$ coordinate system. This gives rise to a partitioning scheme that is substantially different than that generated by DWI (as represented in figure \ref{plot_part}). \\

Therefore, unlike the case with DWI where the subset $\Omega_0$ can be represented as a rectangle in the first quadrant with length $w_0-w_m$ and width $k_0$, the ``rectangle" does not exist in the case for burglary, so the parameterization of $\{w_0,k_0\}$ makes little sense. The more natural parameterization would be perhaps to define the slope $\log\big[ 1+\frac{\tau}{t} \big]$ as a new parameter which essentially governs how ``separated" the three population subsets are. This gives rise to a completely new optimization problem, which may give rise to completely new interesting phase transition behaviors.

\section{Conclusion}

In this work, we developed a model allowing for the maximization of the social welfare function with respect to all penal strategies. The model is general, comprehensive, and practical. \\

The model is general as it uses a graph theoretical approach to explore the dynamics of crimes by representing them as arrows connecting from the offender to the victim. Very few assumptions are made for the arrows, allowing for applicability to a wide range of crimes. To describe a particular class of crime, we simply have to specify the corresponding properties of the arrows (see section \ref{ass}). With the properties specified, the social welfare function can then be constructed explicitly. \\

The model is comprehensive as the social welfare function accounts for all forms of benefits and costs associated with the penal strategy, which include the social benefits of deterring crimes, the social costs of implementing the penal strategy, and the opportunity costs of punishing the offender. This comprehensive collection of social benefits and costs not only allows for an accurate prediction the deterrent effect generated by the penal strategy, but also ensures that the optimal penal strategy captures its full utility for the population. \\

The model is practical in the sense that it can be reduced to an analytically tractable form (see section \ref{solve_opt}) when certain simplifying assumptions are made in the context of the crime of interest (see section \ref{ass}). This can be done for any crime in general (see section \ref{bur}), not just the crime of DWI. This allows the evaluation of the social welfare function to be performed without excessive computational expense, and motivates a deeper understanding of the mechanism behind the interactions of various behavioral traits of the offenders and various factors of penal strategy. Most importantly, the optimal penal strategy can be fully expressed in terms of the distribution parameters of the behavioral traits of the population (see section \ref{dist_tr}), which may be measured empirically (see section \ref{emp}). \\

In this work, we focus on the wealth levels and discount rates of DWI offenders as a case study. In the future, we hope to extend this model to investigate more dynamic crime patterns (see section \ref{dy}), and develop the required mathematical tools for studying this dynamic system. In addition, we hope to model the behavioral traits of offenders more thoroughly by considering other utility parameters beyond the two studied in this work, and conduct an empirical study on a larger sample size with new survey questions formulated to elicit the response of these additional parameters.

\raggedbottom
\pagebreak
\appendix

\section{Properties of the Probability Weighing Function}
\label{pi_prop}

Recall that the probability weighting function is defined in equation \ref{pi} as follow
\begin{equation*}
\pi(p) = \frac{p^{\gamma}}{(p^{\gamma} + (1-p)^{\gamma})^{1/\gamma}},
\end{equation*}
with domain $p\in[0,1]$ and parameter $\gamma\in(0,1)$. It can be easily shown that $\pi(0)=0$ and $\pi(1)=1$, so there are at least two solutions to the equation $\pi(p)=p$. We show that a third solution exists for $0<p<1$. To do so, we first set $\pi(p)=p$
\begin{equation*}
\begin{split}
\pi(p) =& \frac{p^{\gamma}}{(p^{\gamma} + (1-p)^{\gamma})^{1/\gamma}} = p \\
\iff F(p) =& (\frac{\pi(p)}{p})^{\gamma} = \frac{p^{\gamma(\gamma-1)}}{p^{\gamma} + (1-p)^{\gamma}} = 1.
\end{split}
\end{equation*}
Note that the equivalence holds for $p\in [0,1]$. \\

We first evaluate the value (or limit) of $F(p)$ at $p=\{0,\frac{1}{2},1\}$
\begin{equation}
\label{fp}
\lim_{p\to 0}F(p) = \infty \qquad
F(\frac{1}{2}) = 2^{-(1-\gamma)^2}<1 \qquad
F(1) = 1.
\end{equation}
Note that $F(p)$ is a continuous function, so according to the intermediate value theorem, there must be a solution to $F(p)=1$ in $p\in(0,\frac{1}{2})$, which we can denote as $p_0$. We can also show that $F(p)$ is differentiable everywhere and the second derivative must be positive everywhere, or $F''(p)>0$, in $p\in(0,1)$. These properties, combined with equation \ref{fp}, implies that $F(p)$ must have exactly one local minimum in $p\in(0,1)$, and no other critical points. \\

This implies that $p_0$ must be the only solution to $F(p)=1$ in $p\in(0,1)$. Assuming that there are more than one solutions to the equation $F(p)=1$, then $F(p)$ would necessarily have more than two critical points, which creates a contradiction. Furthermore, if we denote the local minimum to be $p^*$, then $\frac{\pi(p)}{p} = \big[ F(p) \big]^{1/\gamma}$ must decrease monotonously for $p\in (0,p^*)$, and increase monotonously for $p\in (p^*,1)$.

\section{Asymptotic Properties of the Partition Curve}
\label{asy}

Given a deterrence strategy targeting $\{w_0,k_0\}$, the pair $\{t,\tau\}$ must satisfy the following constraint
\begin{equation*}
\frac{k_0}{\log[1+\frac{k_0\tau}{1+k_0t}]} = \frac{\pi(p)r}{b-\pi(p)s}.
\end{equation*}
For the sake of clarity, we denote $\psi = \frac{b - \pi(p)s}{\pi(p)r}$, then we can write
\begin{equation}
\label{ap_taut}
\tau = (e^{\psi k_0} - 1)(\frac{1}{k_0} + t).
\end{equation}
We can then solve equation \ref{ap_taut} for $\tau$ and $t$, which allows us to specify the partition curve as appeared in equation \ref{curve}
\begin{equation*}
w = \frac{kf}{r\log\big[ 1+\frac{k\tau}{1+kt} \big]}.
\end{equation*}
Note that in this equation, $k$ is a variable that varies with $w$, instead of being a fixed constant $k_0$. \\

In this form, there are still two main obstacles towards studying the partition curve. The first obstacle is that \ref{ap_taut} is an under-determined equation for $\tau$ and $t$, resulting in infinitely many possible partition curves. The second obstacle is that even if $\{t,\tau\}$ is uniquely specified, the partition curve itself is a transcendental equation, making it analytically intractable. \\

Therefore, we have to approximate the partition curve somehow such that $w$ and $k$ are related polynomially, and the approximate equation does not depend explicitly on $\{t,\tau\}$. For the sake of clarity, we first denote the following function
\begin{equation*}
J(k) = \frac{k}{\log\big[ 1+\frac{k\tau}{1+kt} \big]},
\end{equation*}
with its domain being $k>0$. Then the equation for the partition curve can be written as
\begin{equation*}
w = \frac{f}{r}J(k).
\end{equation*}
We can study the behavior of $J(k)$ in the two limits, $\frac{\tau}{t} \ll 1$ and $\frac{\tau}{t} \gg 1$. We show that in both limits, the approximate form of the partition curve can be fully parameterized by $\{w_0,k_0\}$.

\subsection[small]{Small $\bs{\tau/t}$ Limit}

We first note that $J(k)$ is positive and monotonously increasing in its domain, or $J'(k)>0$ for $k>0$. We study the asymptotic behavior of $J(k)$ by looking at the limits of $J'(k)$ at the two ends of its domain
\begin{equation*}
\lim_{k\to 0}J'(k) = \frac{1}{2} + \frac{t}{\tau}
\qquad
\lim_{k\to\infty}J'(k) = \frac{1}{\log\big[ 1+\frac{\tau}{t} \big]}.
\end{equation*}
This implies that the function $J(k)$ is well behaved in the two limits, meaning that its first derivative approaches a constant. To ensure that the function does not behave unpredictably for intermediate values of $k$, we take the second derivative of the function, which we claim to be negative for all $k$
\begin{equation}
\label{sec_zero}
J''(k) = C_0(k,t,\tau) \Big[ 2k\tau - (2+2kt+k\tau)\log(1+\frac{k\tau}{1+kt})\Big] < 0,
\end{equation}
where $C_0(k,t,\tau)$ is some prefactor that can easily be shown positive. Then to show inequality \ref{sec_zero} to be true, we only have to show that the bracketed term is positive, or
\begin{equation*}
\log(1+\frac{k\tau}{1+kt}) > \frac{k\tau}{1+kt+\frac{k\tau}{2}},
\end{equation*}
which can be derived easily from the following logarithmic inequality
\begin{equation*}
\log(1+x) > \frac{2x}{2+x} \text{ for all } x>0.
\end{equation*}

Now, knowing that $J''(k)<0$, we induce that $J'(k)$ must decrease monotonously from $\lim_{k\to 0}J'(k)= \frac{1}{2}+\frac{t}{\tau}$ to $\lim_{k\to\infty}J'(k) = \big[ \log( 1+\frac{\tau}{t} ) \big]^{-1}$. We can Laurent expand the latter as follow
\begin{equation*}
\lim_{k\to \infty}J'(k) =
\frac{1}{\log\big[ 1+\frac{\tau}{t} \big]} =
\frac{1}{2} + \frac{t}{\tau} + O(\frac{\tau}{t}) =
\lim_{k\to 0}J'(k) + O(\frac{\tau}{t}),
\end{equation*}
where $O(\frac{\tau}{t})$ denotes terms that scale smaller than $\frac{\tau}{t}$ in magnitude. In the limit of $\frac{\tau}{t} \ll 1$, $J'(k)$ is approximately equal at the two limits, and combining this with the fact that $J'(k)$ is a monotonously decreasing function for $k>0$, we see that $J'(k)$ can be approximated as a constant for all $k>0$. Therefore, we can approximate $J(k)$ as a linear function with slope $\frac{1}{2} + \frac{t}{\tau}$. In other words, the partition curve can be approximated as a straight line defined by the following equation
\begin{equation*}
w = \frac{f}{r}J(k) = \frac{f}{r}(\frac{1}{2} + \frac{t}{\tau})k.
\end{equation*}
This is simply the equation for a line that passes through the origin and $\{w_0,k_0\}$, so we can simply the equation to
\begin{equation}
\label{wkwk}
\frac{w}{k} = \frac{w_0}{k_0},
\end{equation}
giving us an expression parameterized only by $\{w_0,k_0\}$.

\subsection[large]{Large $\bs{\tau/tw}$ Limit}

In the limit of $\frac{\tau}{t} \gg 1$, equation \ref{ap_taut} becomes
\begin{equation*}
\frac{\tau}{t} \approx e^{\psi k_0},
\end{equation*}
and the function $J(k)$ can be approximated as
\begin{equation*}
J(k) = \frac{k}{\log\big[ 1+ \frac{k\tau}{1+kt} \big]} \approx \frac{k}{\log\big[ \frac{\tau}{t} \big]} \approx \frac{k}{\psi k_0}.
\end{equation*}
The partition curve is then
\begin{equation*}
w = \frac{f}{r}J(k) = \frac{f}{r}\frac{k}{\psi k_0},
\end{equation*}
which also represents a line that passes through the origin and $\{w_0,k_0\}$. Therefore, the equation for the partition curve is the same as equation \ref{wkwk}, or
\begin{equation*}
\frac{w}{k} = \frac{w_0}{k_0}.
\end{equation*}

\section{Asymptotic Properties of $\bs{J_2}$}
\label{optcost}

We first consider the simple optimization problem where the goal is to find the minimum of $f(t)$ defined as follow
\begin{equation*}
f(t) = \frac{a}{t} + bt,
\end{equation*}
where $t,a,b>0$. It is easy to see that the maximum is attained at $t^* = \sqrt{a/b}$, with the maximum being $f(t^*) = 2\sqrt{ab}$. \\

Recall that the social cost of enforcing imprisonment is given in section \ref{repar} as
\begin{equation}
\label{j2}
\begin{split}
J_2 &\approx \rho\big[ c_0 + \frac{c_t}{mt} + (c_{\tau}-\Lambda l)\tau \big]\,
\Big\{ (1-v^{\alpha})e^{-\kappa_0}+\alpha v^{\alpha}\frac{e^{-\kappa_0}}{\kappa_0} + \alpha\epsilon \frac{e^{-v\kappa_0}}{v\kappa_0} \Big\} \\
&+ \rho\frac{\alpha}{\alpha-1}\big[ gs + (gr+\Lambda b)\tau \big]\,
\Big\{ \big[ (1-v^{\alpha-1})e^{-\kappa_0} + (\alpha-1)v^{\alpha-1}\frac{e^{-\kappa_0}}{\kappa_0} + (\alpha-1)\epsilon\frac{e^{-v\kappa_0}}{v\kappa_0} \big]\Big\}.
\end{split}
\end{equation}
Given a strategy targeting $k_0$, the pair $\{t,\tau\}$ has to satisfy the below constraint (see equation \ref{ap_taut})
\begin{equation*}
\tau = (e^{\psi k_0}-1)(\frac{1}{k_0} + t) \approx e^{\psi \beta \kappa_0},
\end{equation*}
where we assumed that $k_0$ is large and used the denotation $\kappa_0 = \frac{k_0}{\beta}$. Then $J_2$ scales as follow
\begin{equation*}
\begin{split}
J_2 &\sim B(\kappa_0)\frac{A}{t} + B(\kappa_0)\tau \\
&\sim B(\kappa_0)\frac{A}{t} + B(\kappa_0)e^{\psi\beta\kappa_0}t,
\end{split}
\end{equation*}
where $A$ is some constant, and the asymptotic behavior of $B(\kappa_0)$ is governed by
\begin{equation*}
B(\kappa_0) \sim \frac{e^{-v\kappa_0}}{\kappa_0},
\end{equation*}
or simply the slowest decaying term in the large brackets of equation \ref{j2} (note that $v=\frac{w_m}{w_0}<1$). We then see the minimum of $J_2$ over $\{t,\tau\}$ is attained at
\begin{equation*}
\begin{split}
t^* &\sim \sqrt{\frac{B(\kappa_0)A}{B(\kappa_0)e^{\psi\beta\kappa_0}}} \sim \exp\big[ -\frac{\psi\beta}{2}\kappa_0 \big] \\
\tau^* &\sim e^{\psi\beta\kappa_0}t^* = \exp\big[ \frac{\psi\beta}{2}\kappa_0 \big],
\end{split}
\end{equation*}
with the minimal value of $J_2$ being
\begin{equation*}
J_2(\kappa_0,t^*) \sim \sqrt{A B^2(\kappa_0)e^{\psi\beta\kappa_0}} \sim \frac{\exp\big[ (\frac{\psi\beta}{2}-v)\kappa_0 \big]}{\kappa_0}.
\end{equation*}

Therefore, we see that in order for $J_2$ to decay with increasing $\kappa_0$, the following condition must be satisfied
\begin{equation*}
\frac{\psi\beta}{2} - v < 0.
\end{equation*}
Under this condition, we can minimize the total cost of imprisonment by targeting an arbitrarily large $\kappa_0$ value.

\section{Constrained Optimization of Polynomial}
\label{poly}

Consider a polynomial function of the following form
\begin{equation*}
f(x) = x^a(b-cx),
\end{equation*}
where $a,b,c>0$. The first derivative is given by
\begin{equation*}
f'(x) = x^{a-1}\big[ ab-c(a+1)x \big].
\end{equation*}
Setting $f'(x')=0$, we find the critical point to be $x' = \frac{ab}{c(a+1)}$ (where we ignored the trivial $x'=0$). A second derivative test shows $f''(x')<0$, which indicates that it is a local maximum. It can be easily checked that $f'(x)>0$ for $x\in (0,x')$ and $f'(x)<0$ for $x\in (x',+\infty)$. \\

Therefore, for $x\in (0,x_0]$, where $x_0<x'$, the global maximum is attained at $x=x_0$, as $f(x)$ is a monotonously increasing function from $0$ to $x_0$. Similarly, for $x\in [x_0,+\infty)$, where $x_0>x'$, the global maximum is attained at $x=x_0$ as well, as $f(x)$ is a monotonously decreasing function for $x>x_0$. Therefore, given any $x_0\in(0,+\infty)$, the global maximum for $x\in [x_0,+\infty)$ is given by
\begin{equation*}
x^* = \max[ x', x_0 ].
\end{equation*}

\section{Optimization over $\bs{\{v,p\}}$}
\label{two}

Note that in section \ref{final_opt}, the optimization of $J(p,v)$ is essentially solved from a numerical standpoint. However, under a certain condition, there is an analytic expression for the maximum of $J(v,p)$. This is possible when
\begin{equation*}
p_o(1) \geq p_{\min} \iff \frac{bw_m-l}{gsw_m + c_f} \geq \pi^{-1}\big( \frac{b\beta}{2r+s\beta} \big),
\end{equation*}
and the global maximum of $J(v,p)$ is attained at $v=1$ and $p = p_{\min} = \frac{b\beta}{2r+s\beta}$. \\

To see why this is, we first note that $J(v,p_o(v))$ is a monotonously increasing function with respect to $v$ for $v\in[v_{\min},1]$. To see why this is, consider the pair $\{v_1,v_2\}$ where $v_{\min}<v_1<v_2<1$, then the following must be true
\begin{equation*}
J(v_2,p_o(v_2)) > J(v_1,p_o(v_2)) > J(v_1,p_o(v_1)).
\end{equation*}
The first inequality is true because $v_2$ is a global maximum for $J(v,p_o(v_2))$ (see appendix \ref{poly}). The second inequality is true because $J$ is a monotonously decreasing function with respect to $p$ for a fixed $v$ (see equation \ref{jvp}) and $p_o(v)$ is a monotonously decreasing function with respect to $v$ (see equation \ref{vopo}), so
\begin{equation*}
v_2 > v_1 \implies p_o(v_1) > p_o(v_2) \implies J(v,p_o(v_2)) > J(v,p_o(v_1)).
\end{equation*} 
Therefore, $J(v,p_o(v))$ is a monotonously increasing function with respect to $v$, meaning that $J(1,p_o(1))$ is the global maximum on the curve $p_o(v)$. \\

Then for any pair of $p \in [p_{\min},1]$ and $v \in [p_c(v),1]$, we have
\begin{equation*}
J(1,p_{\min}) \geq J(1,p_o(1)) \geq J(v_o(p), p) \geq J(v,p),
\end{equation*}
where the first inequality is a direct result of $p_o(1)\geq p_{\min}$, and the second inequality is due to $J(1,p_o(1))$ being the global maximum on the curve $p_o(v)$, and the last inequality is due to $v_o(p)$ being the global maximum of $J(v,p)$ for a fixed $p$. Therefore, the global maximum of $J$ is attained at $v=1$ and $p=p_{\min}$.

\section{Estimating the Distribution of Discount Rates}
\label{app_dis}

\subsection{Overview of Model}
\label{remo}

Recall that the disutility of imprisonment for an offender is given (in equation \ref{imp_dis}) by
\begin{equation*}
I(w,k,t,\tau) = \frac{rw}{k}\log\big[ 1 + \frac{k\tau}{1+kt} \big].
\end{equation*}
Among $n$ participants in the survey, we determine that the participant $i$ is indifferent between an immediate imprisonment of 2 hours and a delayed imprisonment of $\tau_j$ hours delayed by $t_{ij}$ hours. This gives us the following equation
\begin{equation*}
\begin{split}
& I(w_i,k_i,t_{ij},\tau_j) = I(w_i,k_i,0,2) \\
\implies & t_{ij} = \frac{1}{k_i}(\frac{\tau_j}{2}-1),
\end{split}
\end{equation*}
where $k_i$ is assumed to be the discount rate of the member. \\

The uncertainty of $t_{ij}$ scales with $t_{ij}$ (see section \ref{dist_dis}), so it is necessary to associate $t_{ij}$ with a multiplicative error, giving us the following model
\begin{equation*}
t_{ij}(1+\epsilon_{ij}) = \frac{1}{k_i}(\frac{\tau_j}{2}-1).
\end{equation*}
For the sake of simplicity, we denote $r_{ij} = (\frac{\tau_j}{2}-1)/t_{ij}$, which allows us to write
\begin{equation*}
k_i(1+\epsilon_{ij}) = r_{ij},
\end{equation*}
where the errors $\epsilon_{ij} = \mathcal{N}(0,\sigma_i)$ can be modeled as iid normal RVs with mean $0$ and variance $\sigma_i^2$, which we assume to be different for different participants. \\

The list of imprisonment lengths $\bs{\tau} = \{\tau_1,\tau_2,\tau_3,\tau_4\} = \{ 2.5,4,10,20 \}$ is a fixed vector for all participants, but the number of measurements of $t_{ij}$ made on participant $i$, or $m_i$, depends on the responses of the participant. This is based on the assumption that in a practical scenario, there must a point where the imprisonment is long enough such that the increase in disutility cannot be compensated by any delay in punishment. We can express this as
\begin{equation*}
\exists \tau_0\,[\tau>\tau_0 \implies t=\infty],
\end{equation*}
where $\tau_0$ denotes the ``limit" of imprisonment length that the participant will accept relative to an immediate punishment of 2 hours. Therefore, for $\tau_j > \tau_0$, it makes no sense to measure $t$ anymore as we can expect it to be infinite. We then set the number of measurements for participant $i$ to be
\begin{equation*}
m_i = \min\{ j \,|\, t_{ij}=\infty \}.
\end{equation*}
In other words, the participant will respond with $t_{ij}=\infty$ for $j\geq m_1$, and we can set the data point at $m_i$ as
\begin{equation*}
r_{im_i}=0.
\end{equation*}

Using the $t_{ij}$ measurements, we can estimate $k_i$ for the participant $i$, and using the estimators of $k$ for all the participants, we can construct the distribution of $k$ for the population. We assume that the discount rate $k$ is distributed as a zero-inflated exponential distribution (see equation \ref{pdf2})
\begin{equation*}
f_K(k) = (1-\rho) \delta(k)+\rho\frac{1}{\beta}\exp(-\frac{k}{\beta}),
\end{equation*}
where $\delta$ is the Dirac delta function. This distribution is completely specified by the pair $\{\beta,\rho\}$.

\subsection{Estimating $\mbf{k}$ for a Member}

A natural estimator of $k_i$ would simply be the average of $r_{ij}$ values, or
\begin{equation*}
\hat{k}_i = \frac{1}{m_i} \sum_{j=1}^{m_i} r_{ij},
\end{equation*}
where $m_i$ is the number of measurements for participant $i$. (For the rest of the section, the index $i$ is assumed for $m$, and any sum over $j$ is assumed to be from $1$ to $m_i$.) The estimator $\hat{k}_i$ is unbiased since
\begin{equation*}
\E(\hat{k}_i) = \E\big[ \frac{1}{m}\sum_j k_i(1+\epsilon_{ij}) \big]
= \frac{1}{m}k_i\sum_j\E\big[ 1+\epsilon_{ij} \big] = k_i.
\end{equation*}

For $m>1$, we can estimate the variance of the error $\sigma_i^2$ by evaluating the following expression
\begin{equation*}
\begin{split}
J(\sigma_i^2) = & \frac{1}{m}\sum_j (\frac{r_{ij}}{\hat{k}_i}-1)^2 
= \frac{1}{m} \sum_j \E \Big[ \Big( \frac{1+\epsilon_{ij}}{1+\overline{\epsilon_i}}-1 \Big)^2 \Big] \\
=& \frac{1}{m} \sum_j \E \Big[ \Big( \frac{1+\epsilon_{ij}}{1+\overline{\epsilon_i}}\Big)^2 - 2\frac{1+\epsilon_{ij}}{1+\overline{\epsilon_i}} +1  \Big] \\
=& \frac{1}{m} \E\Big[ \frac{\sum_j(1+\epsilon_{ij})^2}{(1+\overline{\epsilon_i})^2} \Big] - \frac{2}{m} \E\Big[ \frac{m+\sum_j\epsilon_{ij}}{1+\overline{\epsilon_i}} \Big] + 1 
= \E\Big[ \frac{(1+\epsilon_{i1})^2}{(1+\overline{\epsilon_i})^2} \Big] - 1 \\
=& \E\Big[ \Big( \frac{1}{\frac{1}{m} + \frac{1}{m}\frac{\sum_{j\neq 1}(1+\epsilon_{ij})}{1+\epsilon_{i1}}} \Big)^2 \Big] - 1,
\end{split}
\end{equation*}
where we've denoted $\overline{\epsilon_i} = \frac{1}{m}\sum_j \epsilon_{ij}$ as the mean error. We can denote
\begin{equation*}
X = \frac{\frac{1}{m}\sum_{j\neq 1}(1+\epsilon_{ij})}{1+\epsilon_{i1}},
\end{equation*}
as the ratio between two independent normal RVs, with the numerator having mean $\frac{m-1}{m}$ and variance $\frac{m-1}{m^2}\sigma_i^2$, and the denominator having mean $1$ and variance $\sigma_i^2$. The pdf of $X$ can be derived to be (by Jacobian transforming the pdf)
\begin{equation*}
f_X(x) = \frac{m(m-1)(1+mx)}{\sqrt{2\pi}(-1+m+m^2 x^2)^{\frac{3}{2}}\sigma}\exp\Big[ - \frac{(1-m+mx)^2}{2(-1+m+m^2x^2)\sigma^2} \Big].
\end{equation*}
This allows us to numerically evaluate the value of $J(\sigma_i^2)$ for any given pair $\{m,\sigma_i\}$, meaning that we can find $\sigma_i$ after determining the $J(\sigma_i)$ value for the participant\footnote{This can be done, for example, by constructing a look-up table relating a sequence of $\sigma_i$ values (with some small increments between consecutive elements) with the corresponding sequence of $J(\sigma_i^2)$. This is the method that we chose.}. As an example, at $m=4$ and $\sigma \approx 0.2$, the two are related as $J(\sigma_i)\approx 0.77\sigma^2$, which allows us to take
\begin{equation*}
\hat{\sigma_i}^2 \approx \frac{1}{0.77}\frac{1}{m}\sum_j(\frac{r_{ij}}{\hat{k}_i}-1)^2.
\end{equation*}

We can then use $\hat{\sigma}_i$ to compute the standard error of the estimator $\hat{k}_i$
\begin{equation*}
\begin{split}
\E(\hat{k}_i^2) 
&= \frac{1}{m^2} \E\Big\{ \big[ \sum_{j}k_i(1+\epsilon_{ij}) \big]^2 \big\} \\
&= \frac{1}{m^2} k_i^2\sum_{jj'} \E\big[ 1+\epsilon_{ij}+\epsilon_{ij'}+\epsilon_{ij}\epsilon_{ij'} \big] \\
&= \frac{1}{m^2}k_i^2 \sum_{jj'} (1 + \delta_{jj'}\sigma_i^2) 
= k_i^2(1+\frac{\sigma_i^2}{m}) \\
SE(\hat{k}_i) &= \sqrt{\E(\hat{k}_i^2)-\E(\hat{k}_i)^2} = \frac{\hat{k}\hat{\sigma_i}}{\sqrt{m}}.
\end{split}
\end{equation*}
We can then use this standard error to constructed the confidence interval reported in section \ref{dist_dis}. \\

As an important side note, for the special case where $m_i=1$, or when the participant will not choose a punishment of $\tau_1=2.5$ over $2$ hours regardless of the delay, we have $r_{i1}=0$ as the only data point (see section \ref{remo}). To account for this, we simply set the estimator of $k_i$ to be $\hat{k}_i = 0$, with the standard error being $SE(\hat{k}_i)=0$.

\subsection{Estimating $\bs{\{\beta,\rho\}}$ for the Population}

We first begin by introducing a random variable $K'$ which relates to $K$ as follow
\begin{equation*}
K' = \mbf{1}(K \neq 0),
\end{equation*}
where $\mbf{1}$ denotes the indicator function that evaluates to $1$ if its argument is true and $0$ otherwise. Assuming that we are randomly sampling $n$ values of $K$ independently from $f_K(k)$, or $\mbf{K} = \{K_1,K_2,...,K_n\}$, then the number of times that the corresponding $K'$ variables evaluate to 1 can be denoted as 
\begin{equation*}
N' = \sum_{i=1}^n K_i'.
\end{equation*}
We see that $N'$ follows a binomial distribution
\begin{equation*}
Pr(N' = n') = {n \choose n'}\rho^{n'}(1-\rho)^{n'-n'},
\end{equation*}
with the unbiased estimator of $\rho$ being
\begin{equation*}
\hat{\rho} = \frac{N'}{n},
\end{equation*} 
and the standard error of the estimate being
\begin{equation*}
SE(\hat{\rho}) = \sqrt{\frac{\hat{\rho}(1-\hat{\rho})}{n}}.
\end{equation*}

To estimate $\beta$, we focus on the conditional distribution of $f_K(k)$ for non zero $K$, which is simply an exponential distribution with mean $\beta$, or
\begin{equation*}
f_K(k \,|\, K'=1) = \frac{1}{\beta}e^{-\frac{k}{\beta}}.
\end{equation*}
This means that $\beta$ can be estimated from the non-zero values of $k_i$ from the participants. A fairly simple estimator of $\beta$ would be
\begin{equation*}
\hat{\beta} = \big( \sum_{i=1}^n m_i \big)^{-1} \sum_{i=1}^n\sum_{j=1}^{m_i} r_{ij}.
\end{equation*}
The estimator is unbiased since
\begin{equation*}
\begin{split}
\E(\hat{\beta}) &= \big( \sum_{i} m_i \big)^{-1} \sum_i\sum_j\E\big[ k_i(1+\epsilon_{ij}) \big] \\
&= \beta \big( \sum_{i} m_i \big)^{-1} \sum_i\sum_j 1 \\
&= \beta,
\end{split}
\end{equation*}
where the sum of $j$ from $1$ to $m_i$ and the sum of $i$ from $1$ to $n$ is assumed as usual. \\

The variance of this estimator is
\begin{equation*}
\begin{split}
\E(\hat{\beta}_k^2) 
&= \big( \sum_{i} m_i \big)^{-2} \E\Big\{ \big[ \sum_{ij}k_i(1+\epsilon_{ij}) \big]^2 \big\} \\
&= \big( \sum_{i} m_i \big)^{-2} \sum_{ii'}\sum_{jj'} \E\big[ k_ik_{i'} \big] \E\big[ 1 + \epsilon_{ij} + \epsilon_{i'j'} + \epsilon_{ij}\epsilon_{i'j'} \big] \\
&= \big( \sum_{i} m_i \big)^{-2} \sum_{ii'}\sum_{jj'} (\beta^2)(1+\delta_{ii'})(1+\sigma_i^2 \delta_{ii'} \delta_{jj'}) \\
&= \big( \sum_{i} m_i \big)^{-2} \beta^2\sum_{ii'}\sum_{jj'} (1+\delta_{ii'} + 2\sigma_i^2 \delta_{ii'}\delta_{jj'}) \\
&= \beta^2\big( \sum_{i} m_i \big)^{-2}\big[ \big( \sum_{i} m_i \big)^{2} + \sum_i m_i^2 + 2\sum_i m_i\sigma_i^2 \big] \\
&= \beta^2 (1 + \sum_i(\frac{m_i}{\sum_i m_i})^2 + 2\overline{\sigma^2} ),
\end{split}
\end{equation*}
where we denoted
\begin{equation*}
\overline{\sigma^2} = \big( \sum_{i} m_i \big)^{-2} \sum_i m_i\sigma_i^2
\end{equation*}
as the ``weighted" average of the error variances. Therefore, we find the standard error of the estimate to be
\begin{equation*}
\sqrt{\Var(\hat{\mu}_k)}
= \sqrt{\E(\hat{\mu}_k^2) - \big[ \E(\hat{\mu}_k) \big]^2}
= \beta\sqrt{ \sum_i(\frac{m_i}{\sum_i m_i})^2+2\overline{\sigma^2} }.
\end{equation*}

\section{Sample Variance Statistics}
\label{rv_2}

Let $\mbf{X} = \{X_1,...,X_n\}$ be independent normal RVs with zero mean, with the RV $X_i$ having variance $\sigma_i^2$. Denote $\overline{X}$ as the average of the RVs
\begin{equation*}
\overline{X} = \frac{1}{n}\sum_{i=1}^n X_i.
\end{equation*}
We then have
\begin{equation*}
\sum_i X_i^2 = (n-1)S^2 + n \overline{X}^2,
\end{equation*}
where $S^2$ is the sample variance defined as follow
\begin{equation*}
S^2 = \frac{1}{n-1}\sum_i (X_i - \overline{X})^2.
\end{equation*}

Note that $X_i^2\sim \Gamma(\frac{1}{2}, 2\sigma_i^2)$ follows a gamma distribution, with shape $k=\frac{1}{2}$ and rate $\theta = 2\sigma_i^2$. In addition, $n\overline{X}^2\sim \Gamma(\frac{1}{2}, 2\overline{\sigma^2})$ also follows a gamma distribution, with shape $\frac{1}{2}$ and rate $\theta = 2\overline{\sigma^2}$, where $\overline{\sigma^2}$ denotes the average variance of the RVs
\begin{equation*}
\overline{\sigma^2} = \frac{1}{n}\sum_i \sigma_i^2.
\end{equation*}
Since $(n-1)S^2$ and $n\overline{X}^2$ are independent, we have the following relationship for the moment generating functions (noting that the MGF of the addition of two independent RVs is the product of their respective MGFs)
\begin{equation}
\label{mmm}
M_{\sum_i X_i^2}(t) = M_{(n-1)S^2}(t) M_{n\overline{X}^2}(t),
\end{equation}
where
\begin{equation}
\label{mm}
\begin{split}
M_{\sum_i X_i}(t) &= \prod_i M_{X_i}(t) = \prod_i (1-2\sigma_i^2 t)^{-1/2} \\
M_{n\overline{X}^2}(t) &= (1 - 2\overline{\sigma^2}t)^{-1/2}.
\end{split}
\end{equation}
We can then use equation \ref{mmm} to solve for $M_{(n-1)S^2}(t)$ by plugging in the expressions in equation \ref{mm}
\begin{equation*}
M_{(n-1)S^2}(t) = \frac{M_{\sum_i X_i^2}(t)}{M_{n\overline{X}^2}(t)} = \Big[ \prod_i (1-2\sigma_i^2t)^{-1/2} \Big] (1-2\overline{\sigma^2}t)^{1/2}.
\end{equation*}

This allows us to find the first and second moments of the random variable $(n-1)S^2$ to be
\begin{equation*}
\begin{split}
(n-1)\E(S^2) &= M_{(n-1)S^2}'(0) = \big[ \sum_i \sigma_i^2 \big] - \overline{\sigma^2} \\
&= (n-1)\overline{\sigma^2} \\
(n-1)^2\E(S^4) &= M_{(n-1)S^2}''(0) = 3\sum_i \sigma_i^4 + \sum_{i\neq i'}\sigma_i^2\sigma_{i'}^2 -2\overline{\sigma^2}\sum_i \sigma_i^2 - (\overline{\sigma^2})^2 \\
&= 2n\overline{\sigma^4} + (n^2-2n-1)(\overline{\sigma^2})^2,
\end{split}
\end{equation*}
which allows us to find the variance of $S^2$
\begin{equation*}
\begin{split}
\Var(S^2) &= \E(S^4) - \big[ \E(S^2)\big]^2 \\
&= \frac{1}{(n-1)^2}\big\{ 2n\overline{\sigma^4} + (n^2-2n-1)(\overline{\sigma^2})^2 - (n-1)^2(\overline{\sigma^2})^2 \big\} \\
&= 2\frac{n\overline{\sigma^4}-(\overline{\sigma^2})^2}{(n-1)^2}.
\end{split}
\end{equation*}
This result will be useful when we study the standard error of the variance estimator of the distribution of $\gamma$ (probability weighing factor) in appendix \ref{app_prob} (the following section).

\section{Estimating Probability Weighing Factor}
\label{app_prob}

The participant $i$ is indifferent between spending $B_{ij}$ amount of money for hiring a dedicated driver and committing DWI under a $p_j$ chance of apprehension. We then set the utility function to be a normal RV of mean $0$, or $\epsilon_i = \mathcal{N}(0,\sigma_i)$
\begin{equation*}
B_{ij} - \pi(p_j,\gamma_i) (S_i + D_i) = \epsilon_{i},
\end{equation*}
where $D_i$ is the disutiltiy of the punishment and $S_i$ is the stigma associated with apprehension (both of which we do not know the values to). Note that $S_i+D_i$ are $\gamma_i$ are two unknown constants for participant $i$, so they constitute the two regression coefficients in this analysis. \\

This presents a non-linear regression problem, where the goal is to minimize the following for each $i$
\begin{equation*}
\sum_{j=1}^{m} \big[ B_{ij} - (S_i+D_i)\pi(p_j,\gamma_i) \big]^2,
\end{equation*}
over $\gamma_i$ and $S_i+D_i$. We can find the estimator $\hat{\gamma}_i$ and its confidence interval using standard statistical analysis packages, such as the non-linear regression toolbox in MATLAB. We denote the standard error of $\gamma_i$ as $\sigma_{\gamma_i}$. \\

If we assume that among the population, $\gamma$ is distributed normally with mean $\mu_{\gamma}$ and variance $\sigma_{\gamma}^2$, then we see that the estimator $\hat{\gamma_i}$ is a normal RV with mean $\mu_{\gamma}$ and variance $\sigma_{\gamma}^2 + \sigma_{\gamma_i}^2$. We then easily see that the unbiased estimator of $\mu_{\gamma}$ and its standard error is given by
\begin{equation*}
\hat{\mu}_{\gamma} = \frac{1}{n}\sum_i \hat{\gamma}_i 
\qquad
SE(\hat{\mu_{\gamma}}) = \sqrt{\frac{1}{n}} \sqrt{\sigma_{\gamma}^2 + \overline{\sigma_{\gamma_i}^2}},
\end{equation*}
where $\overline{\sigma_{\gamma_i}^2}$ denotes the average standard variance of $\hat{\gamma}_i$, or
\begin{equation*}
\overline{\sigma_{\gamma_i}^2} = \frac{1}{n}\sum_i \sigma_{\gamma_i}^2.
\end{equation*}
Furthermore, using the results from appendix \ref{rv_2}, we can find the unbiased estimator of $\sigma_{\gamma}^2$ (the variance of the distribution of $\gamma$) and its standard error to be
\begin{equation*}
\begin{split}
\hat{\sigma_{\gamma}^2} &= \frac{1}{n-1} \sum_i (\hat{\gamma}_i - \overline{\hat{\gamma}})^2 - \overline{\sigma_{\gamma_i}^2}
\\
SE(\hat{\sigma_{\gamma}^2}) &= \frac{\sqrt{2}}{n-1}\sqrt{(n-1)\sigma_{\gamma}^4 + 2(n-1)\sigma_{\gamma}^2 \overline{\sigma_{\gamma_i}^2} + n\overline{\sigma_{\gamma_i}^4} - (\overline{\sigma_{\gamma_i}^2})^2}.
\end{split}
\end{equation*}
This allows us to construct the confidence interval for $\hat{\mu_{\gamma}}$ and $\hat{\sigma_{\gamma}^2}$, the estimators of mean and variance of the distribution of $\gamma$ among the population. Note that we can easily derive the estimator and the standard error for $\sigma_{\gamma}$ using the Delta method (if we assume that $\hat{\sigma_{\gamma}^2}$ is approximately normal, which is true in the limit of large $n$), and this gives us
\begin{equation*}
\hat{\sigma_{\gamma}} = \sqrt{\hat{\sigma_{\gamma}^2}}
\qquad
SE(\hat{\sigma_{\gamma}}) = \frac{1}{2\hat{\sigma_{\gamma}}}SE(\hat{\sigma_{\gamma}^2}).
\end{equation*}

\raggedbottom
\pagebreak

\end{document}